\documentclass[aps,prl,twocolumn,superscriptaddress]{revtex4}

\usepackage{amssymb}
\usepackage{natbib}
\usepackage{afterpage}

\usepackage{multirow}
\usepackage{bm}
\usepackage{amsmath}
\usepackage{graphicx}
\usepackage{epstopdf}
\usepackage{subfigure}
\usepackage{natbib}
\usepackage{epsfig}
\usepackage{amsfonts}
\usepackage{mathrsfs}
\usepackage{braket}
\usepackage{tabularx}
\usepackage[toc,page,title,titletoc,header]{appendix}
\usepackage{hyperref}
\hypersetup{colorlinks,allcolors=blue}

\usepackage{dsfont,amsthm,amsbsy}

\usepackage{fancyhdr}
\usepackage{lineno}
\begin{document}

\title{Uncovering the spin ordering in magic-angle graphene via edge state equilibration}


\author{Jesse C. Hoke}
\affiliation{Department of Physics, Stanford University, Stanford, CA 94305, USA}
\affiliation{Geballe Laboratory for Advanced Materials, Stanford, CA 94305, USA}
\affiliation{Stanford Institute for Materials and Energy Sciences, SLAC National Accelerator Laboratory, Menlo Park, CA 94025, USA}

\author{Yifan Li}
\affiliation{Department of Physics, Stanford University, Stanford, CA 94305, USA}
\affiliation{Geballe Laboratory for Advanced Materials, Stanford, CA 94305, USA}
\affiliation{Stanford Institute for Materials and Energy Sciences, SLAC National Accelerator Laboratory, Menlo Park, CA 94025, USA}

\author{Julian May-Mann}
\affiliation{Department of Physics, Stanford University, Stanford, CA 94305, USA}
\affiliation{Department of Physics and Institute for Condensed Matter Theory, University of Illinois at Urbana-Champaign, Urbana, IL 61801, USA}

\author{Kenji Watanabe}
\affiliation{Research Center for Electronic and Optical Materials,
National Institute for Materials Science, 1-1 Namiki, Tsukuba 305-0044, Japan}

\author{Takashi Taniguchi}
\affiliation{Research Center for Materials Nanoarchitectonics,
National Institute for Materials Science, 1-1 Namiki, Tsukuba 305-0044, Japan}

\author{Barry Bradlyn}
\affiliation{Department of Physics and Institute for Condensed Matter Theory, University of Illinois at Urbana-Champaign, Urbana, IL 61801, USA}

\author{Taylor L. Hughes}
\affiliation{Department of Physics and Institute for Condensed Matter Theory, University of Illinois at Urbana-Champaign, Urbana, IL 61801, USA}

\author{Benjamin E. Feldman}
\email{bef@stanford.edu}
\affiliation{Department of Physics, Stanford University, Stanford, CA 94305, USA}
\affiliation{Geballe Laboratory for Advanced Materials, Stanford, CA 94305, USA}
\affiliation{Stanford Institute for Materials and Energy Sciences, SLAC National Accelerator Laboratory, Menlo Park, CA 94025, USA}


\begin{abstract}

\textbf{ABSTRACT:}
The flat bands in magic-angle twisted bilayer graphene (MATBG) provide an especially rich arena to investigate interaction-driven ground states. While progress has been made in identifying the correlated insulators and their excitations at commensurate moiré filling factors, the spin–valley polarizations of the topological states that emerge at high magnetic field remain unknown. Here we introduce a technique based on twist-decoupled van der Waals layers that enables measurement of their electronic band structure and--by studying the backscattering between counter-propagating edge states--the determination of the relative spin polarization of their edge modes. We find that the symmetry-broken quantum Hall states that extend from the charge neutrality point in MATBG are spin unpolarized at even integer filling factors. The measurements also indicate that the correlated Chern insulator emerging from half filling of the flat valence band is spin unpolarized and suggest that its conduction band counterpart may be spin polarized. 
    
\end{abstract}

\maketitle
\section{Introduction}

\begin{figure*}[t!]
    \renewcommand{\thefigure}{\arabic{figure}}
    \centering
    \includegraphics[width=1.99\columnwidth]{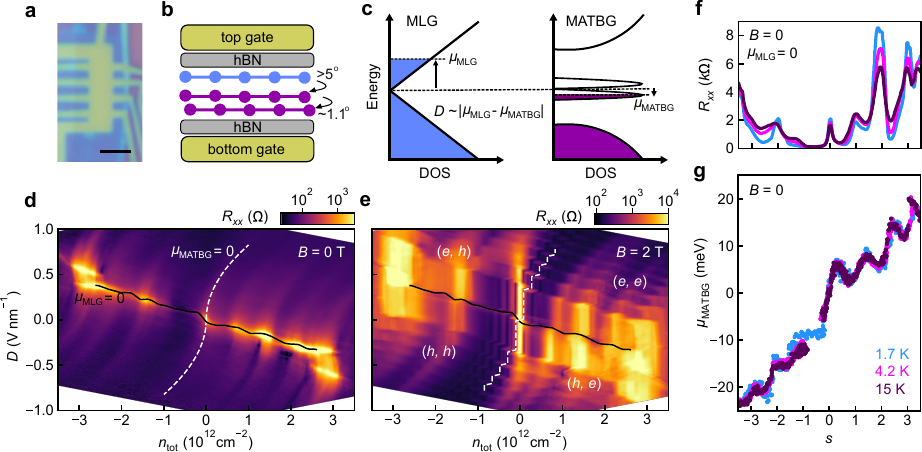}
    \caption{\textbf{Twist-decoupled monolayer graphene (MLG) and magic-angle twisted bilayer graphene (MATBG)}. \textbf{a}, Optical image of the device. The scale bar is 2 $\mu$m. \textbf{b}, Schematic of the device structure and interlayer angles. The twisted trilayer graphene is encapsulated in hexagonal boron nitride (hBN) and has graphite top and bottom gates. \textbf{c}, Band diagram of the combined MLG-MATBG system. The displacement field $D$ modifies the energies of states in each subsystem and therefore tunes the relative chemical potential $\mu_i$ of each subsystem $i$ at fixed total carrier density $n_{\mathrm{tot}}$. \textbf{d}, \textbf{e}, Longitudinal resistance $R_{xx}$ as a function of $n_{\mathrm{tot}}$ and $D$, at zero magnetic field $B$ and at $B =$ 2 T, respectively. Black solid (white dashed) lines denote where the MLG (MATBG) is at its charge neutrality point (CNP). Parentheses indicate which carrier types are present in the MLG and MATBG, respectively: $e$ indicates electrons and $h$ indicates holes. \textbf{f}, $R_{xx}$ as a function of moiré filling factor $s$ at $B = 0$ and at various temperatures $T$ where the MLG is at its CNP (solid black curve in \textbf{d}). \textbf{g}, $\mu_{\mathrm{MATBG}}$ as a function of $s$ at $B = 0$, as extracted from \textbf{d} and analogous data at other temperatures.
} 
    \label{fig:Fig1}
\end{figure*}

The relative twist angle between adjacent van der Waals layers provides a powerful tuning knob to control electronic properties. In the limit of large interlayer twist, the misalignment leads to a mismatch in the momentum and/or internal quantum degrees of freedom of low-energy states in each layer, resulting in effectively decoupled electronic systems~\cite{sanchez-yamagishi_quantum_2012, sanchez-yamagishi_helical_2017, rickhaus_electronic_2020, rickhaus_correlated_2021, mrenca-kolasinska_quantum_2022, shi_bilayer_2022, Kim_robust_2022}. This decoupling can be sufficiently pronounced to realize independently tunable quantum Hall bilayers that support artificial quantum spin Hall states~\cite{sanchez-yamagishi_helical_2017} or excitonic condensation~\cite{shi_bilayer_2022, Kim_robust_2022}. In the opposite regime of low twist angle, a moiré superlattice develops, and can lead to extremely flat electronic bands with prominent electron-electron interaction effects. The archetypal low-twist example is magic-angle twisted bilayer graphene (MATBG)~\cite{cao_correlated_2018, cao_unconventional_2018, bistritzer_moire_2011}, which has been shown to support symmetry-broken quantum Hall states~\cite{cao_unconventional_2018, choi_correlation-driven_2021, yankowitz_tuning_2019, lu_superconductors_2019} as well as correlated Chern insulators (ChIs) at high magnetic fields~\cite{yu_correlated_2022, das_symmetry-broken_2021, saito_hofstadter_2021, nuckolls_strongly_2020, wu_chern_2021, choi_correlation-driven_2021, Tomarken_2019, park_flavour_2021, pierce_unconventional_2021, xie_fractional_2021}. However, a full understanding of the nature of these states, including their spin and valley polarization, has so far remained elusive.

Combining large and small interlayer twists in a single device provides an approach to probe microscopic details of correlated ground states in moiré systems~\cite{Andrei_2020, Balents_2020, Mak_2022}. Such a device would yield electronically decoupled flat and dispersive bands which can be used to interrogate each other. In some ways, this is reminiscent of other two-dimensional heterostructures which host bands of differing character. One notable example is mirror-symmetric magic-angle twisted trilayer graphene (MATTG) and its multilayer generalizations~\cite{park_tunable_2021, hao_electric_2021, liu_isospin_2022, park_robust_2022, zhang_ascendance_2021, khalaf_magic_2019, Shen_2022}, which can be decomposed into flat MATBG-like bands that coexist with more dispersive bands. However, these bands hybridize at non-zero displacement field, whereas a twist-decoupled architecture provides fully independent bands. This enables control over the relative filling of light and heavy carriers, including in a bipolar (electron-hole) regime. Crucially, in a perpendicular magnetic field, such a device can realize a quantum Hall bilayer with co- or counter-propagating edge modes. Because the inter-edge mode coupling depends on their respective internal degrees of freedom~\cite{sanchez-yamagishi_helical_2017,amet_2014}, the effects of edge backscattering on transport can be used to identify spin/valley flavor polarization of the flat moiré bands.

Here we report transport measurements of a dual-gated, twisted trilayer graphene device that realizes electrically decoupled MATBG and monolayer graphene (MLG) subsystems. By tracking features in the resistance as a function of carrier density and displacement field, we demonstrate independently tunable flat and dispersive bands and show that transport measurements can be used to simultaneously determine the thermodynamic density of states in each subsystem. Furthermore, in the regime of counter-propagating MLG and MATBG edge modes in a magnetic field, we use longitudinal and non-local resistance measurements to infer the spin order within the MATBG subsystem--both for symmetry-broken quantum Hall states emanating from the charge neutrality point (CNP), and for the primary sequence of ChIs. Our work clarifies the microscopic ordering of correlated states in MATBG and demonstrates a powerful generic method to probe internal quantum degrees of freedom in two-dimensional electron systems.

\section{Results}

\subsection{Twist-decoupled Flat and Dispersive Bands}

\begin{figure*}[t!]
    \renewcommand{\thefigure}{\arabic{figure}}
    \centering
    \includegraphics[width=1.99\columnwidth]{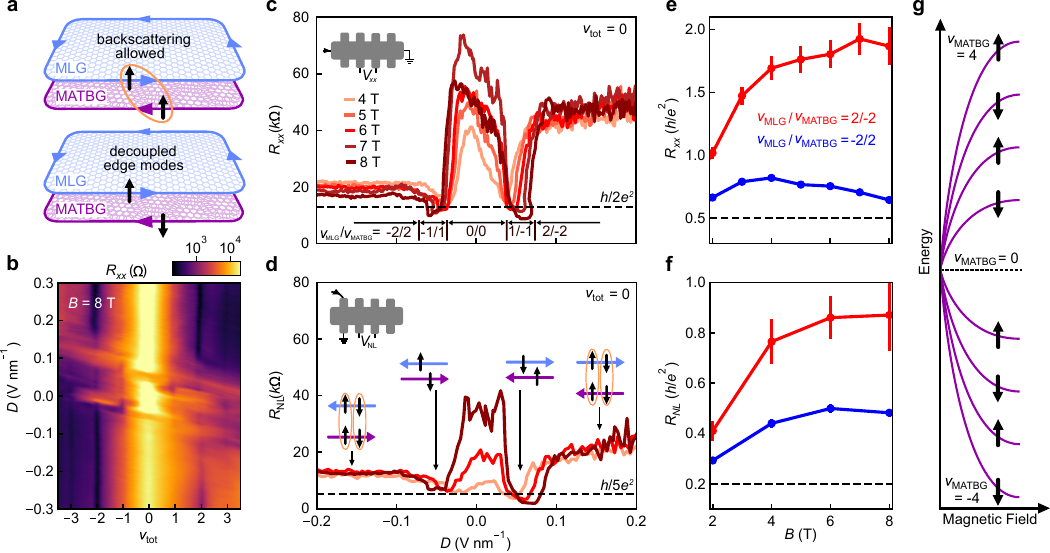}
    \caption{\textbf{Spin polarization of MATBG quantum Hall states near the CNP}. \textbf{a}, Schematic illustration of two possible scenarios for a single pair of counter-propagating edge modes. If the spins of each edge mode are aligned (top), backscattering is allowed (orange circle). Backscattering is suppressed when the spins are anti-aligned (bottom), leading to quantum spin Hall-like behavior with $R_{xx} = h/2e^{2}$. \textbf{b}, $R_{xx}$ as a function of the total filling factor $\nu_{\mathrm{tot}} = \nu_{\mathrm{MLG}} + \nu_{\mathrm{MATBG}}$ and $D$ at $B = 8$ T. \textbf{c}, \textbf{d} $R_{xx}$ and $R_{\mathrm{NL}}$, respectively measured in the configurations shown in the top left insets, as a function of $D$ when $\nu_{\mathrm{tot}} = 0$. Filling factors of each subsystem for each regime of $D$ are indicated in the bottom inset of \textbf{c}. Insets in \textbf{d} schematically represent the inferred relative spin orientations (black arrows) of edge modes in MLG (blue arrows) and MATBG (purple arrows), with orange circles indicating backscattering between a given pair. \textbf{e}, \textbf{f}, $R_{xx}$ and $R_\mathrm{NL}$ for $\nu_{\mathrm{MATBG}} = \pm 2 / \mp 2$ (red and blue, respectively) averaged over $0.1 < |D| < 0.25$ V nm$^{-1}$. Error bars correspond to one standard deviation. The straight lines connecting data points are guides for the eye. \textbf{g}, Schematic diagram of CNP MATBG Landau levels (LLs) and their spin characters. Gaps between LLs are depicted schematically and do not represent experimentally measured field dependence.
} 
    \label{fig:Fig2}
\end{figure*}

An optical image of the device is shown in Fig.~\ref{fig:Fig1}\textbf{a}, with a side view of individual layers schematically illustrated in Fig.~\ref{fig:Fig1}\textbf{b}. As we demonstrate below, the bottom two graphene layers have a twist of $1.11^{\circ}$ and display behavior consistent with typical MATBG samples, while the topmost graphene layer is electrically decoupled because of the larger interlayer twist of approximately $5-6^{\circ}$ (see Methods). The whole device is encapsulated in hexagonal boron nitride (hBN) and has graphite top and bottom gates. This dual gated structure allows us to independently tune the total carrier density $n_{\mathrm{tot}} = (C_{b} V_{b} + C_{t} V_{t})/e$ and applied displacement field $D = (C_{t} V_{t} - C_{b} V_{b})/(2 \epsilon_{0})$, where $C_{b(t)}$ and $V_{b(t)}$ are the capacitance and voltage of the bottom (top) gate, $e$ is the electron charge, and $\epsilon_{0}$ is the vacuum permittivity. The applied displacement field shifts the relative energies of states in each subsystem and therefore controls how the total carrier density is distributed between them (Fig.~\ref{fig:Fig1}\textbf{c}).

We first describe electronic transport through the device at zero magnetic field. The longitudinal resistance $R_{xx}$ is largest along a curve at low/moderate $D$, with multiple fainter, S-shaped resistive features extending outward, i.e. approximately transverse to it (Fig.~\ref{fig:Fig1}\textbf{d}). This phenomenology arises from electronic transport in parallel through the MLG and MATBG subsystems. Specifically, the strongly resistive behavior occurs when the MLG is at its CNP (solid black line in Fig.~\ref{fig:Fig1}\textbf{d}). Relatively higher peaks in $R_{xx}$ along this curve reflect insulating states in MATBG. Analogously, when the carrier density in MATBG is fixed to an insulating state, $R_{xx}$ remains elevated even as the carrier density in the MLG is adjusted. This leads to the resistive S-shaped curves (such as the dashed white line in Fig.~\ref{fig:Fig1}\textbf{d}; see discussion below).

The peaks in $R_{xx}$ centered near $n_{\mathrm{tot}} = \pm{2.8 \times 10^{12}}$ cm$^{-2}$ correspond to the single-particle superlattice gaps at moiré filling factor (number of electrons per unit cell) $s = \pm{4}$. From these densities, we extract a twist angle of $\theta = 1.11^{\circ}$ between the bottom two layers, and similar measurements using different contact pairs show that there is little twist angle disorder in these two layers (Supplementary Fig. S1). Intermediate resistance peaks are also present at $s = 0$, $1$, $\pm{2}$, and $3$ (Fig.~\ref{fig:Fig1}\textbf{d},\textbf{f}).
These peaks are consistent with the correlated insulators that have been previously observed in MATBG \cite{cao_correlated_2018, yankowitz_tuning_2019, lu_superconductors_2019, saito_independent_2020, stepanov_untying_2020, liu_tuning_2021, yu_spin_2022, morissette_electron_2022, nuckolls2023quantum}
and they persist as the MLG is doped away from its CNP (Supplementary Fig. S2).
At higher temperatures, another peak develops near $s = -1$ (Supplementary Fig. S3), matching prior reports of a Pomeranchuk-like effect in MATBG \cite{saito_isospin_2021, rozen_entropic_2021}. 

Our characterization demonstrates the ability to independently tune the carrier density in each subsystem, and hence shows that the subsystems are effectively decoupled. This further allows the MLG to act as a thermodynamic sensor for the MATBG, similar to schemes in which a sensing graphene flake is isolated by a thin hBN spacer from the target sample~\cite{park_flavour_2021, saito_isospin_2021, liu_isospin_2022, kim_direct_2012}. By tracking the resistive maxima when the MLG is at its CNP, and using a model that accounts for screening of electric fields by each layer (Supplementary Note 2), we extract the MATBG chemical potential $\mu_\mathrm{MATBG}$ (Fig.~\ref{fig:Fig1}\textbf{g}). We find a total change of chemical potential across the flat bands of $\delta \mu \approx 40$ meV, with non-monotonic dependence on filling that matches previous reports of a sawtooth in inverse compressibility \cite{yu_correlated_2022, pierce_unconventional_2021, saito_isospin_2021, rozen_entropic_2021, zondiner_cascade_2020}. Similarly, we can determine the MLG chemical potential as a function of its carrier density $\mu_{\mathrm{MLG}}(n_{\mathrm{MLG}})$ by fitting to the S-shaped resistive features in Fig.~\ref{fig:Fig1}\textbf{d}, which occur at fixed $s$ in MATBG (Supplementary Note 2). These match the scaling $\mu_{\mathrm{MLG}} \propto \mathrm{sgn}(n_\mathrm{MLG}) |n_{\mathrm{MLG}}|^{1/2}$ that is expected for the Dirac dispersion of graphene. We observe similar behavior in a second trilayer device, where MLG-like states are decoupled from a bilayer graphene moiré system with a $1.3^{\circ}$ twist angle (Supplementary Fig. S4), suggesting this is a generic phenomenon that is widely applicable in multilayer heterostructures.

Electronic decoupling is also evident when we apply a perpendicular magnetic field $B$, where the energy spectrum of MLG consists of Landau levels (LLs), and a Hofstadter butterfly spectrum develops in MATBG. Figure~\ref{fig:Fig1}\textbf{e} shows $R_{xx}$ as a function of $n_{\mathrm{tot}}$ and $D$ at $B = 2$ T, revealing staircase-like patterns which reflect crossings of the MLG LLs and MATBG states (Hall resistance $R_{xy}$ is plotted in Supplementary Fig. S5). Vertical features at constant $n_{\mathrm{tot}}$ occur when the MLG is in a quantum Hall state; their extent (in $D$) is proportional to the size of the gap between LLs. As the displacement field tunes the relative energies of states in each subsystem, transitions occur when graphene LLs are populated or emptied. These cause each feature associated with a MATBG state to shift horizontally in density by the amount needed to fill a fourfold degenerate LL, $n_{\mathrm{LL}} = 4eB/h$, where $h$ is Planck’s constant and the factor of four accounts for the spin and valley degrees of freedom (e.g., see dashed white line in Fig.~\ref{fig:Fig1}\textbf{e}).

\subsection{Quantum Hall Edge State Equilibration}

In a magnetic field, the decoupled MLG and MATBG realize a quantum Hall bilayer in which either carrier type (electron or hole) can be stabilized in either subsystem. This results in co- (counter-)propagating edge modes when the respective carrier types are the same (different). Additionally, because the device is etched into a Hall bar after stacking, the edges of MLG and MATBG are perfectly aligned. Crucially, in the counter-propagating regime, the measured resistance encodes information about the efficiency of scattering between the edge modes in each subsystem (Supplementary Note 3), which depends on their internal quantum degrees of freedom. We expect that atomic scale roughness at the etched edge of the device enables large momentum transfer, and therefore anticipate efficient coupling irrespective of valley (in MLG and MATBG) and moiré valley (in MATBG). However, assuming the absence of magnetic disorder, edge states having different spins should remain decoupled, whereas those with the same spin can backscatter and exhibit increased longitudinal resistance (Fig.~\ref{fig:Fig2}\textbf{a}). Probing $R_{xx}$ therefore allows us to deduce the relative spin polarization of edge states in MLG and MATBG.

We first focus at low carrier density and high magnetic field, where the behavior of each subsystem  $i$ is well described by quantum Hall states having filling factors $\nu_i = n_ih/eB$ emanating from their respective CNPs. A sharp peak in $R_{xx}$ emerges at combined filling factor $\nu_{\mathrm{tot}} = 0$, flanked by several quantum Hall states at other integer $\nu_{\mathrm{tot}}$ (Fig.~\ref{fig:Fig2}\textbf{b}). These features exhibit a series of $D$-field tuned transitions as the relative filling of MLG and MATBG changes. The data encompass MLG states with $|\nu_\mathrm{MLG}| \leq 2$. Importantly, prior work has shown that MLG edge modes at $\nu_\mathrm{MLG} = \pm1$ have opposite spin and valley quantum numbers, whereas those at $\nu_\mathrm{MLG} = \pm2$ are spin unpolarized~\cite{amet_2014}. Combining this information with the measured resistance enables us to determine the spin polarization of the MATBG quantum Hall states with $|\nu_{\mathrm{MATBG}}| \leq 4$.

When $\nu_{\mathrm{tot}} = 0$, MLG and MATBG have equal and opposite filling, and $R_{xx}$ approaches different values depending on the number of counter-propagating edge states (Fig.~\ref{fig:Fig2}\textbf{c}). At $D = 0$, each subsystem is in an insulating, $\nu=0$ symmetry-broken state. Here, no bulk conduction or edge modes are anticipated, and we observe a large resistance. Near $|D|\approx 0.05$~V/nm, $\nu_{\mathrm{MLG}}/\nu_{\mathrm{MATBG}} = \pm 1 / \mp 1$, and $R_{xx}$ reaches a minimum near $h/2e^{2}$ (Fig.~\ref{fig:Fig2}\textbf{c}). This phenomenology can be explained by a pair of counter-propagating edge modes with opposite spins, analogous to helical edge modes observed in large-angle twisted bilayer graphene \cite{sanchez-yamagishi_helical_2017}. This interpretation is further corroborated by similar behavior in another contact pair (Supplementary Note 4), and measurements of non-local resistance $R_{\mathrm{NL}}$ (Fig.~\ref{fig:Fig2}\textbf{d}). Indeed, the pronounced non-local resistance signal at $\nu_{\mathrm{MLG}}/\nu_{\mathrm{MATBG}} = \pm 1 / \mp 1$ indicates that transport is dominated by edge modes (see Supplementary Note 5 for a discussion of bulk effects). This is corroborated by the value of $R_\mathrm{NL}$, which is suppressed toward $h/5e^2$, the quantized value predicted from the Landauer--Büttiker formula for counter-propagating edge states in this contact configuration (Supplementary Note 3). We therefore conclude that similar to MLG, MATBG has a filled spin down (up) electron- (hole-)like LL at $\nu_\mathrm{MATBG} = 1 (-1)$.

Beyond $|D|\approx 0.08$ V/nm, where $\nu_{\mathrm{MLG}}/\nu_{\mathrm{MATBG}} = \pm 2 / \mp 2$, we observe larger resistances $R_{xx} > h/2e^{2}$ and $R_\mathrm{NL} > h/5e^{2}$ (Fig.~\ref{fig:Fig2}\textbf{c},\textbf{d}). This suggests that backscattering occurs for both pairs of edge modes: if both MATBG edge states had identical spin, one counter-propagating pair would remain decoupled and would lead to quantized resistance $R_{xx} = h/2e^2$ and $R_\mathrm{NL} = h/5e^2$ (Supplementary Note 3). A resistance above this value, as well as the large increase in resistance relative to $\nu_{\mathrm{MLG}}/\nu_{\mathrm{MATBG}} = \pm 1 / \mp 1$, therefore both indicate that the edge states at $\nu_{\mathrm{MATBG}} = \pm 2$ are spin unpolarized (see Supplementary Note 4-5 for additional measurements and discussion of alternative interpretations which we rule out as unlikely). There is some asymmetry in the measured $R_{xx}$ depending on the sign of $D$; it comparatively less pronounced in $R_{\mathrm{NL}}$. Since $R_{\mathrm{NL}}$ is inherently a probe of edge conduction, this suggests the observed asymmetry in $R_{xx}$ originates from additional bulk current contributions, which may arise due to an electron-hole asymmetry in the strengths of different symmetry-broken states (see Supplementary Note 5).
Based on the above observations, we deduce the spin polarization of the edge modes of the MATBG LLs emanating from its CNP as illustrated in Fig.~\ref{fig:Fig2}\textbf{g}. 

\begin{figure*}[t!]
    \renewcommand{\thefigure}{\arabic{figure}}
    \centering
    \includegraphics[width=1.9\columnwidth]{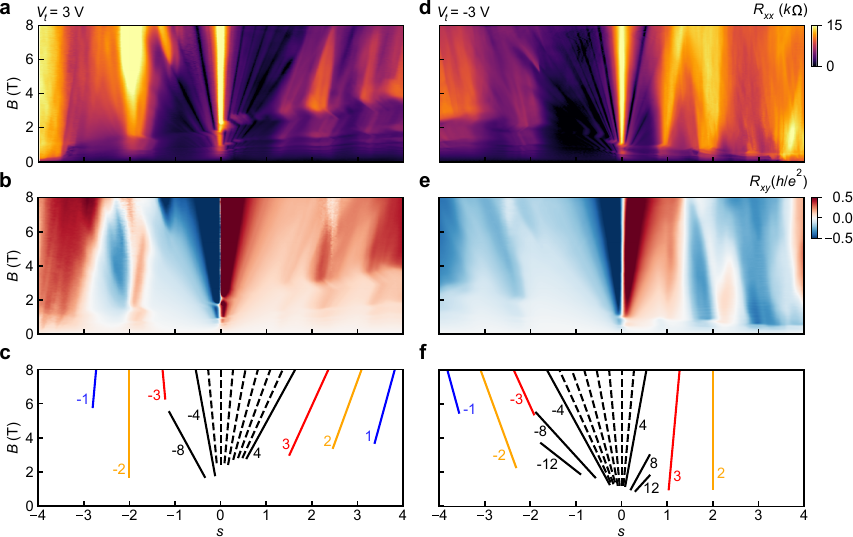}
    \caption{\textbf{Landau fans demonstrating correlated Chern Insulators (ChIs)}. \textbf{a}, \textbf{b}, $R_{xx}$ and $R_{xy}$ as a function of $s$ and $B$ at fixed top gate voltage $V_t = 3$ V. \textbf{c}, Wannier diagram indicating the strongest quantum Hall and ChI states determined from \textbf{a} and \textbf{b}.  The Chern numbers $t$ of the MATBG states are labeled. At high fields, the total Chern numbers of each state are offset by $2$ because $\nu_{\mathrm{MLG}} = 2$. Black, red, orange, and blue lines correspond to states with zero-field intercepts $s=0$, $s=|1|$,  $s=|2|$, and $s=|3|$, respectively. For states with $s=0$, $t\equiv\nu_\mathrm{MATBG}$. Black dashed lines label the MATBG symmetry-broken quantum Hall states $-4 < \nu_\mathrm{MATBG} < 4$. \textbf{d}, \textbf{e}, \textbf{f}, Same as \textbf{a}, \textbf{b}, \textbf{c}, but for $V_t = -3$, where $\nu_{\mathrm{MLG}} = -2$ at high fields. Data collected at $T \approx 300$ mK.  
} 
    \label{fig:Fig3}
\end{figure*}

\begin{figure*}[t!]
    \renewcommand{\thefigure}{\arabic{figure}}
    \centering
    \includegraphics[width=1.6\columnwidth]{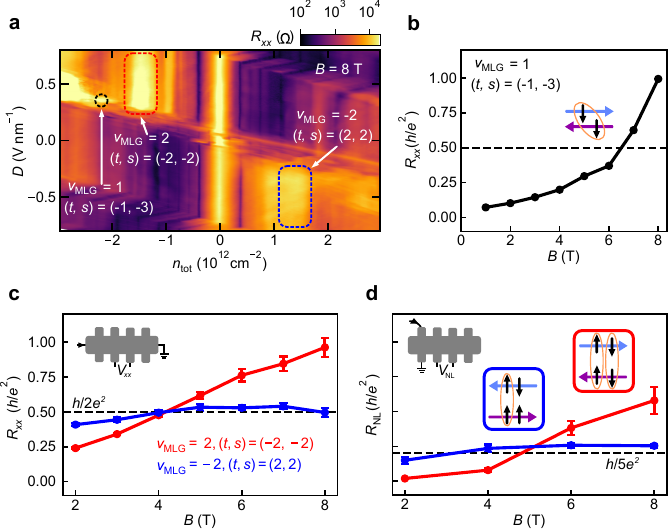}
    \caption{\textbf{Spin polarization of the ChIs in MATBG}. \textbf{a}, $R_{xx}$ as a function of $n_\mathrm{tot}$ and $D$ at $B = 8$ T (see Supplementary Fig. S8 for equivalent map in a non-local contact configuration). Black dashed circle: $\nu_{\mathrm{MLG}} = 1, (t, s) = (-1, -3)$. Red dashed box: $\nu_{\mathrm{MLG}} = 2, (t, s) = (-2, -2)$. 
    Blue dashed box: $\nu_{\mathrm{MLG}} = -2, (t, s) = (2, 2)$. 
    \textbf{b}, $R_{xx}$ for the $\nu_{\mathrm{MLG}} = 1, (t, s) = (-1, -3)$ state as a function of $B$. \textbf{c}, \textbf{d}, $R_{xx}$ and $R_{\mathrm{NL}}$, respectively, measured in the configurations shown in the top left insets, for $\nu_{\mathrm{MLG}} = \pm 2, (t, s) = (\mp 2, \mp 2)$ states (red and blue, respectively) as a function of $B$. Data are averaged over $0.325 < |D| < 0.525$ V nm$^{-1}$. Error bars correspond to one standard deviation. Insets in \textbf{d} schematically represent the inferred relative spin orientations (black arrows) of edge modes in MLG (blue arrows) and MATBG (purple arrows), with orange circles indicating backscattering between a given pair. The straight lines connecting data points are guides for the eye.
} 
    \label{fig:Fig4}
\end{figure*}

\subsection{Addressing Spin Polarization of the Chern Insulators}

In addition to symmetry-broken quantum Hall states emerging from the CNP, ChIs extrapolating to nonzero $s$ are evident in Landau fan measurements of $R_{xx}$ and $R_{xy}$ at fixed top gate voltages of $\pm3$ V (Fig.~\ref{fig:Fig3}). At these values, the MLG filling factor is $\nu_\mathrm{MLG} = \pm2$, respectively, at high fields. Consequently, both the Chern number of the primary sequence of quantum Hall states in MATBG (black lines in Fig.~\ref{fig:Fig3}\textbf{c,f}) emerging from $s=0,$ and the ChIs (colored lines) are offset by $\pm2$. After accounting for this shift, the ChIs that we observe are consistent with the primary sequence $|t + s| = 4$ commonly reported in MATBG, where $t$ is the Chern number of the MATBG subsystem~\cite{yu_correlated_2022, das_symmetry-broken_2021, saito_hofstadter_2021, nuckolls_strongly_2020, wu_chern_2021, choi_correlation-driven_2021, Tomarken_2019, park_flavour_2021}. Below, we focus primarily on the $(t, s) = (\pm2, \pm2)$ ChIs, which exhibit near-zero $R_{xx}$ and quantized $R_{xy}$ in the co-propagating regime (Supplementary Fig. S6). Here, ChI edge mode chirality is determined by the sign of $t$: states with $t>0$ $(t<0)$ have electron- (hole-)like edge modes.

Tuning into the bipolar (electron-hole) regime, allows us to realize counter-propagating edge modes from the MATBG ChIs and the MLG quantum Hall states. We apply the edge state equilibration analysis to determine the spin polarization of the ChIs in MATBG. For the $(t, s) = (-1, -3$) ChI, we find a sharp resistive feature that occurs only when $\nu_{\mathrm{MLG}} = 1$ (Fig.~\ref{fig:Fig4}\textbf{a-b}), i.e. when there is one pair of counter-propagating edge states. The resistance grows with increasing $B$ and reaches values significantly larger than $h/2e^{2}$ (Fig.~\ref{fig:Fig4}\textbf{b}). This indicates strong backscattering between edge modes, and hence that both have the same spin (inset, Fig.~\ref{fig:Fig4}\textbf{b}). We conclude that the first flavor to occupy the MATBG Hofstadter subbands (see Supplementary Note 6) is spin down, consistent with expectations based on the Zeeman effect.

A resistive state also occurs when $(t, s) = (-2, -2)$ and $\nu_{\mathrm{MLG}} = 2$ (Fig.~\ref{fig:Fig4}\textbf{a}). We observe $R_{xx} > h/2e^{2}$ that grows with increasing $B$ (Fig.~\ref{fig:Fig4}\textbf{c} and Supplementary Fig. S7), indicating efficient backscattering between both pairs of counter-propagating edge modes. We obtain consistent results from both the non-local resistance (Fig.~\ref{fig:Fig4}\textbf{d}) and $R_{xx}$ measurements of a second contact pair (Supplementary Note 4). We therefore conclude that the $(-2, -2)$ ChI in MATBG is spin unpolarized (red inset, Fig.~\ref{fig:Fig4}\textbf{d}).

In contrast, we observe more moderate resistance for the $(t, s) = (2, 2)$ ChI in MATBG when $\nu_{\mathrm{MLG}} = -2$ (Fig.~\ref{fig:Fig4}\textbf{a}). In measurements of $R_{xx}$ ($R_{\mathrm{NL}}$) at fixed $B$, the resistance of this state saturates near $h/2e^{2}$ ($h/5e^{2}$) at high $B$ (Fig.~\ref{fig:Fig4}\textbf{c}, \textbf{d}), with similar near-quantized $R_{xx}$ in a Landau fan measurement (Supplementary Fig. S7). Together, these results demonstrate that there is only partial coupling between edge modes. The data are consistent with one pair of decoupled, counter-propagating edge modes, and another pair having allowed backscattering. This would naturally arise if the $(t, s) = (2, 2)$ ChI in MATBG is spin polarized (blue inset, Fig.~\ref{fig:Fig4}d). The data therefore suggest a spin polarized ground state may be favored (see Supplementary Notes 5-6 for further discussion).

\section{Discussion}

The observed spin orderings of both the quantum Hall states and the ChIs clarify the microscopic interactions and relative strengths of different symmetry breaking terms in MATBG. Near charge neutrality, spin unpolarized states are favored at $\nu_{\mathrm{MATBG}} = \pm 2$ for all measured magnetic fields $B > 2$ T (Fig.~\ref{fig:Fig2}\textbf{e-f}). This is counter to expectations based on both the conventional Hofstadter subband model (see Supplementary Note 6) and Zeeman considerations. Specifically, moiré valley splitting~\cite{yu_correlated_2022}, which arises in the presence of $M_y$ symmetry breaking, or some other mixing between Hofstadter subbands is necessary to produce a spin unpolarized state at $\nu_{\mathrm{MATBG}} = \pm 2$ (see Supplementary Note 6). Moreover, even in the presence of moiré valley splitting, the Zeeman effect would favor spin polarization at $\nu_{\mathrm{MATBG}} = \pm 2$; our observations therefore indicate that exchange interactions dominate over Zeeman splitting throughout the measured field range and favor spin unpolarized states.

Very recent theoretical work~\cite{wang2023theory} suggests there is a crossover between spin polarized ChIs favored by the Zeeman effect at high magnetic field and a partially spin unpolarized intervalley coherent state favored at low magnetic field, with the former predicted to dominate at experimentally relevant fields. Our results for the $(t, s) = (2, 2)$ ChI are consistent with the high field prediction, but the spin unpolarized states we observe at $(t, s) = (-2, -2)$ are not. This discrepancy likely reflects electron-hole asymmetry in MATBG and/or atomic scale relaxation of the lattice, which are neglected in the theoretical model. The calculations indicate close competition between different ground states, so including these effects will alter quantitative predictions and could even lead to qualitatively different ground states, as observed experimentally. Our work provides an important benchmark for future theoretical considerations, demonstrating the importance of these terms, that distinct spin ordering can occur for electron and hole doping, and that antiferromagnetic exchange contributions can be comparable to or larger in magnitude than Zeeman splitting. 

In conclusion, we have realized a twisted graphene multilayer consisting of electrically decoupled MATBG and MLG subsystems. Even though the layers are in contact, we demonstrated that a twist-decoupled architecture provides a method to extract thermodynamic properties and probe internal quantum degrees of freedom through edge state equilibration. Looking forward, we anticipate its extension to other van der Waals materials, including to recently discovered systems that exhibit fractional quantum anomalous Hall states~\cite{cai2023signatures, zeng2023thermodynamic, park2023observation, xu2023fractional, lu2023fractional}. This device geometry also represents the most extreme limit of dielectric screening of interactions \cite{saito_independent_2020, stepanov_untying_2020, liu_tuning_2021} in which a tunable screening layer is immediately adjacent to the system of interest. More generally, it provides a natural arena to explore Kondo lattices \cite{kumar_gate-tunable_2022, doniach_kondo_1977} with independently tunable densities of itinerant electrons and local moments, as well as an opportunity to study Coulomb drag between adjacent layers \cite{narozhny_coulomb_2016}.


\section{Methods}
\subsection{Device fabrication}
The MATBG-MLG stack was fabricated using standard dry transfer techniques with poly (bisphenol A carbonate)/polydimethylsiloxane (PC/PDMS) transfer slides~\cite{wang_one-dimensional_2013, zomer_fast_2014}. A monolayer graphene flake was cut into 3 pieces with a conductive AFM tip in contact mode. An exfoliated hBN flake (26.5 nm) was used to sequentially pick up each section at the desired twist angle before placing it on top of a prefabricated stack of few-layer graphite and hBN (27 nm). Finally, an additional few-layer graphite flake was added and served as the top gate. The stack was subsequently patterned with standard electron beam lithography techniques followed by etching to form a Hall bar geometry and metallization to form edge contacts~\cite{wang_one-dimensional_2013}.

\subsection{Transport measurements}
Transport measurements were conducted at cryogenic temperatures ($1.7$ K unless otherwise stated) using standard lock-in techniques with a current bias of 5 - 20 nA at 17.777 Hz. Because edge contacts are made to the etched sample, they simultaneously make electrical contact to all three graphene layers, and electronic transport through the device reflects parallel transport through both MATBG and MLG subsystems. The longitudinal and transverse resistance are symmetrized and anti-symmetrized in a magnetic field, respectively.

\subsection{Determination of twist angle}
The twist angle $\theta$ between the pair of layers that form MATBG is determined by the superlattice carrier density $n_{s} = 4/A \approx 8\theta^{2}/\sqrt{3}a^{2}$ when the MLG is at charge neutrality. Here, $A$ is the superlattice unit cell area and $a = 0.246$ nm is the MLG lattice constant. The twist angle can also be independently confirmed by fitting Chern insulators in a Landau fan measurement using the Streda formula $dn/dB = C_{\mathrm{tot}}/\Phi_{0}$, which have intercepts at zero magnetic field at integers $s = 4n/n_{s}$. Both methods yield a consistent value of $\theta = 1.11^{o} \pm 0.05^{o}$, where the quoted uncertainty reflects the width of the $s = \pm 4$ resistive peaks. The capacitances $C_{b}$ ($C_{t}$) between the bottom (top) gate and sample are accurately determined based on the slopes of features in Landau fans taken at constant bottom (top) gate voltages $V_{b}$ ($V_{t}$). This also results in vertical features (Fig.~\ref{fig:Fig1}\textbf{e}, for example) when data are plotted as a function of total carrier density $n_{\mathrm{tot}} = C_{t}V_{t}/e + C_{b}V_{b}/e$ and displacement field $D =(C_{t}V_{t}/e - C_{b}V_{b}/e)/(2\epsilon_{0})$, where $e$ is the electron charge, and $\epsilon_{0}$ is the vacuum permittivity. 

The relative angle between the MLG and MATBG subsystems is estimated based on the angle between the AFM-cut edges of the top and middle graphene layers. An optical image of the three graphene (and top hBN) layers on the PC/PDMS stamp during the stacking process (before deposition) is shown in Supplementary Fig. S9. From the image, we identify a twist angle between the top and middle graphene layers of about $5.5^{\circ} \pm 0.5^{\circ} $.

\section{Data availability}
Data that support the findings in this study are available at https://doi.org/10.5281/zenodo.11044381. Additional datasets generated and/or analyzed during the current study are available from the corresponding author upon request.

\section{Code availability}
The codes that support the findings of this study are available from the corresponding authors upon request.

\section{References
}
\bibliography{references.bib}

\section{Acknowledgements}
We thank Pablo Jarillo-Herrero, Steve Kivelson, Yves Kwan, Sid Parameswaran, B. Andrei Bernevig, Oskar Vafek, Xiaoyu Wang, David Goldhaber-Gordon, and Aaron Sharpe for helpful discussions. This work was supported by the QSQM, an Energy Frontier Research Center funded by the U.S. Department of Energy (DOE), Office of Science, Basic Energy Sciences (BES), under Award \# DE-SC0021238. K.W. and T.T. acknowledge support from the JSPS KAKENHI (Grant Numbers 20H00354 and 23H02052) and World Premier International Research Center Initiative (WPI), MEXT, Japan. J.C.H. acknowledges support from the Stanford Q-FARM Quantum Science and Engineering Fellowship. Part of this work was performed at the Stanford Nano Shared Facilities (SNSF), supported by the National Science Foundation under award ECCS-2026822.

\section{Author contribution}
J.C.H. fabricated the devices. J.C.H. and Y.L. conducted transport measurements. B.E.F supervised the project. J.M.M., B.B., and T.L.H. provided theoretical support. K.W. and T.T provided hBN crystals. All authors contributed to analysis and writing of the manuscript. 

\section{Competing interests}
The authors declare no competing interest.


\end{document}


\title{Supplementary Notes for: \\ Uncovering the spin ordering in magic-angle graphene via edge state equilibration}

\author{Jesse C. Hoke}
\affiliation{Department of Physics, Stanford University, Stanford, CA 94305, USA}
\affiliation{Geballe Laboratory for Advanced Materials, Stanford, CA 94305, USA}
\affiliation{Stanford Institute for Materials and Energy Sciences, SLAC National Accelerator Laboratory, Menlo Park, CA 94025, USA}

\author{Yifan Li}
\affiliation{Department of Physics, Stanford University, Stanford, CA 94305, USA}
\affiliation{Geballe Laboratory for Advanced Materials, Stanford, CA 94305, USA}
\affiliation{Stanford Institute for Materials and Energy Sciences, SLAC National Accelerator Laboratory, Menlo Park, CA 94025, USA}

\author{Julian May-Mann}
\affiliation{Department of Physics, Stanford University, Stanford, CA 94305, USA}
\affiliation{Department of Physics and Institute for Condensed Matter Theory, University of Illinois at Urbana-Champaign, Urbana, IL 61801, USA}

\author{Kenji Watanabe}
\affiliation{Research Center for Electronic and Optical Materials,
National Institute for Materials Science, 1-1 Namiki, Tsukuba 305-0044, Japan}

\author{Takashi Taniguchi}
\affiliation{Research Center for Materials Nanoarchitectonics,
National Institute for Materials Science, 1-1 Namiki, Tsukuba 305-0044, Japan}

\author{Barry Bradlyn}
\affiliation{Department of Physics and Institute for Condensed Matter Theory, University of Illinois at Urbana-Champaign, Urbana, IL 61801, USA}

\author{Taylor L. Hughes}
\affiliation{Department of Physics and Institute for Condensed Matter Theory, University of Illinois at Urbana-Champaign, Urbana, IL 61801, USA}

\author{Benjamin E. Feldman}
\email{bef@stanford.edu}
\affiliation{Department of Physics, Stanford University, Stanford, CA 94305, USA}
\affiliation{Geballe Laboratory for Advanced Materials, Stanford, CA 94305, USA}
\affiliation{Stanford Institute for Materials and Energy Sciences, SLAC National Accelerator Laboratory, Menlo Park, CA 94025, USA}

\maketitle

\section{1. Additional Figures}

\begin{figure*}[h!]
    \renewcommand{\thefigure}{S\arabic{figure}}
    \centering
    \includegraphics[width=0.9\columnwidth]{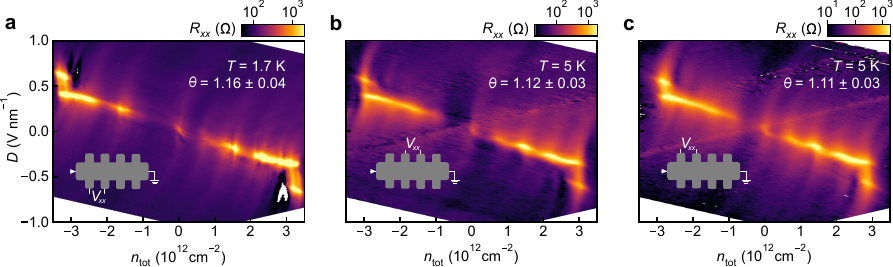}
    \caption{\textbf{Sample homogeneity}.
    Longitudinal resistance $R_{xx}$ as a function of total carrier density $n_\mathrm{tot}$ and displacement field $D$ for three different contact pairs separate from the one discussed in the main text.
} 
    \label{fig:contacts}
\end{figure*}

\begin{figure*}[h!]
    \renewcommand{\thefigure}{S\arabic{figure}}
    \centering
    \includegraphics[width=0.8\columnwidth]{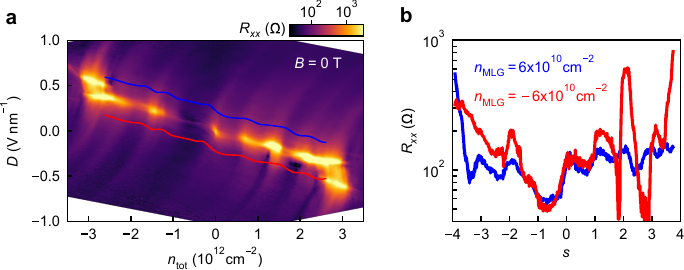}
    \caption{\textbf{Resistance of MATBG at non-zero density of the monolayer graphene}. \textbf{a}, $R_{xx}$ as a function of $n_\mathrm{tot}$ and $D$ at temperature $T = 1.7$ K and magnetic field $B = 0$ T. Blue and red curves represent where the monolayer graphene is at fixed $n_{\mathrm{MLG}} = \pm 6 \times 10^{10}$ cm$^{-2}$, respectively. \textbf{b}, $R_{xx}$ as a function of moiré filling factor $s$ along the blue and red curves in \textbf{a}.
} 
    \label{fig:Rxx_nonzero_nMLG}
\end{figure*}

\newpage

\begin{figure*}[h!]
    \renewcommand{\thefigure}{S\arabic{figure}}
    \centering
    \includegraphics[width=0.65\columnwidth]{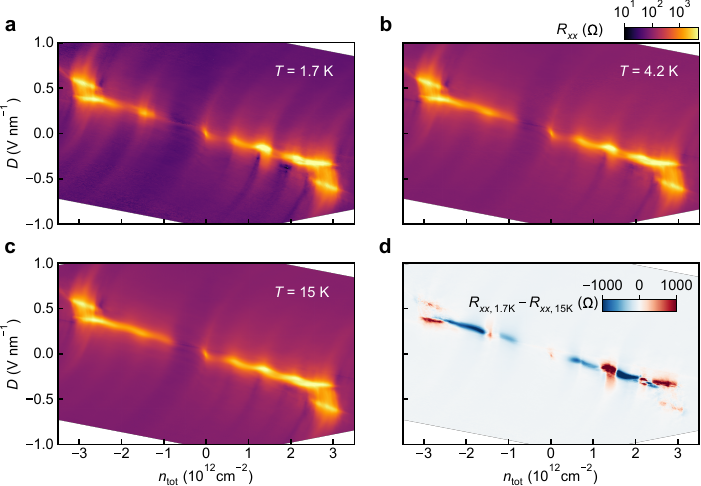}
    \caption{\textbf{Temperature dependence at zero magnetic field}. \textbf{a}, \textbf{b}, \textbf{c}, $R_{xx}$ as a function of $n_\mathrm{tot}$ and $D$ at temperature $T$ = 1.7 K, 4.2 K and 15 K, respectively. The appearance of a resistive peak near moiré filling factor $s$ = -1 at higher temperatures reflects a Pomeranchuk-like effect in magic-angle twisted bilayer graphene (MATBG). \textbf{d}, $R_{xx, 1.7 \mathrm{K}} - R_{xx, 15 \mathrm{K}}$ as a function of $n_\mathrm{tot}$ and $D$, revealing insulating temperature dependence at $s = \pm 2, 3,$ and $\pm 4$.
} 
    \label{fig:Tdependence}
\end{figure*}

\begin{figure*}[h!]
    \renewcommand{\thefigure}{S\arabic{figure}}
    \centering
    \includegraphics[width=0.45\columnwidth]{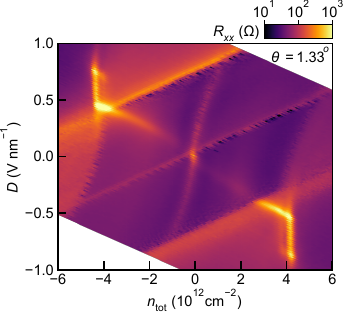}
    \caption{\textbf{Transport of second device}. $R_{xx}$ as a function of $n_\mathrm{tot}$ and $D$ at $T$ = 5 K for a second twist-decoupled device. Based on transport (see Methods), we estimate a twist angle of $\theta \approx 1.33^{\circ}$ between the near-aligned layers that form its moiré subsystem. An analysis of an optical image (see Methods) of the stack indicates the twist angle between the moiré subsystem and the monolayer graphene subsystem is approximately $15^{\circ}$. Resistive peaks appear at $s = 0$ and $\pm 4$ when the decoupled graphene monolayer is at its charge neutrality point. An S-shaped resistive feature near $n_{tot} = 0$ extends towards higher displacement fields as the graphene carrier density is adjusted while maintaining a fixed density of the moiré subsystem. This feature also displays the expected Dirac scaling $\mu_{\mathrm{MLG}} \propto \mathrm{sgn}(n_\mathrm{MLG}) |n_{\mathrm{MLG}}|^{1/2}$. This phenomenology is consistent with the primary device reported in the main text. The straight diagonal features (which are independent of $V_t$) are due to parallel transport in a portion of the device that is only bottom gated. 
} 
    \label{fig:device2}
\end{figure*}

\newpage

\begin{figure*}[h!]
    \renewcommand{\thefigure}{S\arabic{figure}}
    \centering
    \includegraphics[width=0.45\columnwidth]{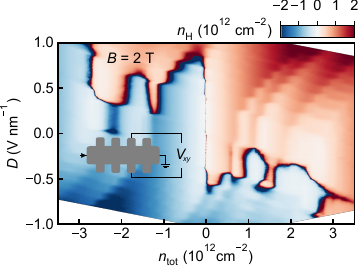}
    \caption{\textbf{Hall density at $B = 2$ T}. Hall density $n_\mathrm{H}$ as a function of $n_\mathrm{tot}$ and $D$ at $B$ = 2 T. The Hall density qualitatively matches what is expected for the combined system, with resets in the effective MATBG carrier density occurring near certain integer $s$.}

    \label{fig:Hall}
\end{figure*}

\begin{figure*}[h!]
    \renewcommand{\thefigure}{S\arabic{figure}}
    \centering
    \includegraphics[width=0.6\columnwidth]{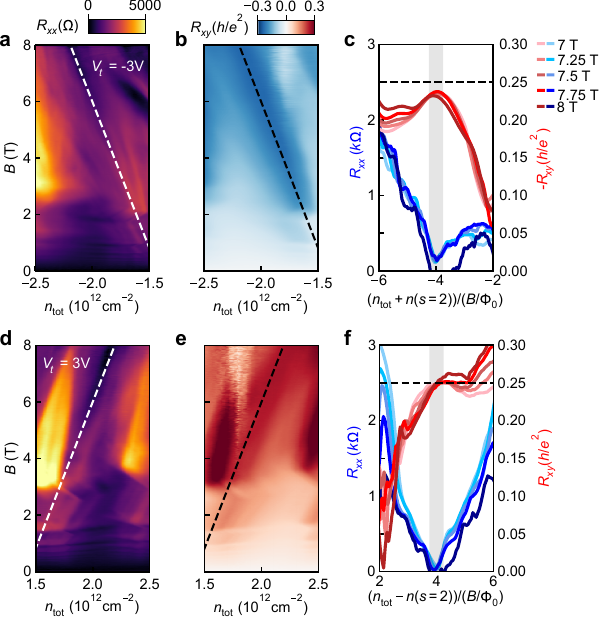}
    \caption{\textbf{Landau fan measurement of even Chern insulators in the co-propagating regime}. \textbf{a}, \textbf{b}, $R_{xx}$ and $R_{xy}$ as a function of $n_{\mathrm{tot}}$ and $B$ at fixed top gate voltage $V_t = -3$ V near the $(t, s) = (-2, -2)$ Chern insulator (ChI), where $t$ is the Chern number. \textbf{c}, Line cuts of $R_{xx}$ and $R_{xy}$ showing well developed minima near zero $R_{xx}$ and near-quantized $R_{xy}$ for the $\nu_{\mathrm{MLG}}=-2$/$(t, s) = (-2, -2)$ state. \textbf{d}, \textbf{e}, \textbf{f}, Same as \textbf{a}, \textbf{b}, \textbf{c}, but for the $\nu_{\mathrm{MLG}}=2$/$(t, s) = (2, 2)$ state when $V_t = 3$. Data are identical to that presented in Fig. 3, but zoom-ins and line cuts of the relevant features are shown here for clarity. Data collected at $T \approx 300$ mK.
} 
    \label{fig:Landau_Chern_Zoom_Co}
\end{figure*}

\newpage

\begin{figure*}[h!]
    \renewcommand{\thefigure}{S\arabic{figure}}
    \centering
    \includegraphics[width=0.9\columnwidth]{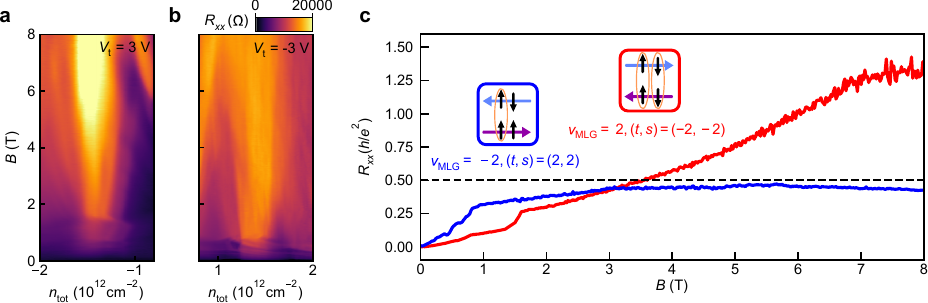}
    \caption{\textbf{Landau fan measurement of even Chern insulators in the counter-propogating regime}. \textbf{a}, \textbf{b}, $R_{xx}$ as a function of $n_{\mathrm{tot}}$ and $B$ at fixed $V_t = 3$ V and $V_t = -3$ V, respectively, near the $\nu_{\mathrm{MLG}}=\pm2$/$(t, s) = (\mp2, \mp2)$ states. \textbf{c}, $R_{xx}$ as a function $B$ for the  $\nu_{\mathrm{MLG}}=2$/$(t, s) = (-2, -2)$ (red) and $\nu_{\mathrm{MLG}}=-2$/$(t, s) = (2, 2)$ (blue) states. Data are identical to that presented in Fig. 3, but zoom-ins and line cuts of the relevant features are shown here for clarity. Data collected at $T \approx 300$ mK.
} 
    \label{fig:Landau_Chern_Zoom_Counter}
\end{figure*}

\begin{figure*}[h!]
    \renewcommand{\thefigure}{S\arabic{figure}}
    \centering
    \includegraphics[width=0.5\columnwidth]{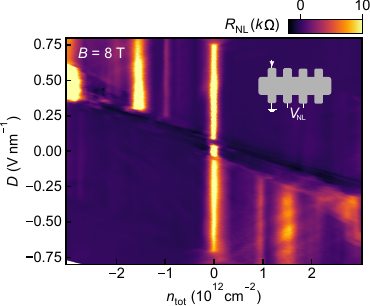}
    \caption{\textbf{Non-local transport} Non-local resistance $R_\mathrm{NL}$ measured using the contact configuration shown in the inset as a function of $n_\mathrm{tot}$ and $D$ at $B = 8$ T.
} 
    \label{fig:non_local}
\end{figure*}

\newpage

\begin{figure*}[h!]
    \renewcommand{\thefigure}{S\arabic{figure}}
    \centering
    \includegraphics[width=0.7\columnwidth]{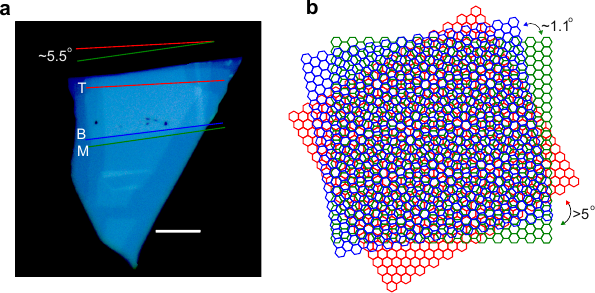}
    \caption{\textbf{Twist angle between graphene layers}. \textbf{a}, Optical image of the trilayer stack after being picked up with an hBN flake on the polymer stamp but before deposition onto the bottom gate. Red, green, and blue lines are drawn on top of the AFM cut edges of the individual top, middle, and bottom monolayer flakes, respectively. From this optical image we estimate a twist angle between the top and middle layer of $\sim 5-6^{\circ}$. The white scale bar is $10 \ \mu$m. \textbf{b}, Schematic representation of the 3 graphene layers and their relative twists. Interlayer angles are exaggerated in the image for clarity.
} 
    \label{fig:angles}
\end{figure*}

\section{2. Chemical potential extraction and interlayer capacitance estimate}

\begin{figure*}[b!]
    \renewcommand{\thefigure}{S\arabic{figure}}
    \centering
    \includegraphics[width=1\columnwidth]{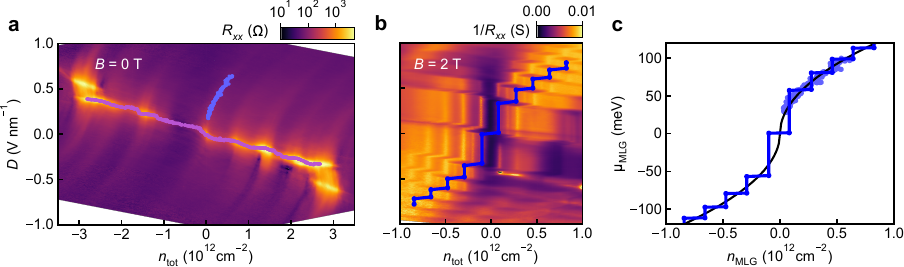}
    \caption{\textbf{Extraction of the MLG and MATBG chemical potentials, and interlayer capacitance}. \textbf{a},  Longitudinal resistance $R_{xx}$ as a function of total carrier density $n_\mathrm{tot}$ and displacement field $D$ at zero magnetic field $B$ with the extracted $R_{xx}$ maxima corresponding to where $\mu_\mathrm{MATBG} = 0$ (light blue) and $\mu_\mathrm{MLG} = 0$ (light purple). \textbf{b}, $1/R_{xx}$ as a function of $n_\mathrm{tot}$ and $D$ at $B = 2$ T. The dark blue curve tracks the maxima of $1/R_{xx}$ where $\mu_\mathrm{MATBG} = 0$. \textbf{c}, Extracted MLG chemical potentials $\mu_{\mathrm{MLG}}$ at $B = 0$ (light blue) and $B = 2$ T (dark blue) using the points in \textbf{a} and \textbf{b}. A least squares fit of the data to the theoretically predicted functional form of the graphene chemical potential at $B = 0$ is shown in black.
} 
    \label{fig:chem_pot}
\end{figure*}

To extract the chemical potentials of the MLG and MATBG subsystems, we use a model based on the screening of electric fields similar to that in Supplementary Refs.~\cite{kim_direct_2012, park_flavour_2021, mrenca-kolasinska_quantum_2022, liu_isospin_2022}. Although there is no hBN spacer between the MLG and the MATBG in our device, the large twist angle between these subsystems electronically decouples them so that a similar analysis applies and we can model the two subsystems with an interlayer capacitance $C_i$. This model yields the following equations for the densities $n_i$ and chemical potentials $\mu_i$ of a given subsystem $i$ when the other remains at charge neutrality:
\begin{equation}
n_\mathrm{MLG} = C_t V_t / e + (C_t + C_i)C_b V_b / e C_i,    
\end{equation}
\begin{equation}
n_\mathrm{MATBG} = C_b V_b / e + (C_b + C_i) C_t V_t /e C_i,  
\end{equation}
\begin{equation}
\mu_\mathrm{MLG} = -e C_b V_b / C_i,    
\end{equation}
\begin{equation}
\mu_\mathrm{MATBG} = -e C_t V_t/C_i 
\end{equation}
Here, $e$ is the electron charge, and $C_{b(t)}$ and $V_{b(t)}$ are respectively the geometric capacitance and voltage applied to the bottom (top) gate. In the above equations, $C_i$ is an unknown. However, because $\mu_\mathrm{MLG} = \mathrm{sgn}(n_\mathrm{MLG}) \hbar v_F (\pi |n_\mathrm{MLG}|)^{1/2}$, $C_i$ can be used as a fitting parameter to Supplementary equation 3 and experimentally determined ($\hbar$ is the reduced Planck’s constant and we take the graphene Fermi velocity $v_F = 1.12 \times 10^6$ m/s). This value for $C_i$ can then be used in Supplementary equation 4 to obtain $\mu_\mathrm{MATBG}(n_\mathrm{MATBG})$. To find $C_i$, we use the locations of the maxima in longitudinal resistance $R_{xx}$ along $\mu_\mathrm{MATBG} = 0$ (light blue points, Fig.~\ref{fig:chem_pot}\textbf{a}) and use a least squares fitting procedure to match them to the theoretically expected $\mu_\mathrm{MLG}(n_\mathrm{MLG})$ (black curve, Fig.~\ref{fig:chem_pot}\textbf{c}), yielding a best fit $C_i = 5.24 \pm 0.52$  $\mu \mathrm{F}$ $\mathrm{cm}^{-2}$. This is consistent with previous estimates in other twisted graphene systems, falling between the values extracted for twisted bilayer graphene and twisted double bilayer graphene~\cite{sanchez-yamagishi_quantum_2012, rickhaus_electronic_2020, Rickhaus_2019}; while we cannot rule out that there is some hybridization between layers which renormalizes the Fermi velocity of the MLG subsystem~\cite{Uri_2023}, the agreement supports the validity of assuming a bare graphene Fermi velocity $v_F = 1.12 \times 10^6$ m/s.

A similar procedure can be used to extract $\mu_\mathrm{MLG}(n_\mathrm{MLG})$ at nonzero perpendicular magnetic field $B$. In Fig.~\ref{fig:chem_pot}\textbf{b} we plot $1/R_{xx}$ at $B = 2$ T, where clear Landau level (LL) crossings are observed. Overlaid in dark blue are lines which track $n_\mathrm{MATBG} = 0$. The resulting MLG chemical potential extracted from the points in Fig.~\ref{fig:chem_pot}\textbf{b} is displayed in Fig.~\ref{fig:chem_pot}\textbf{c}.

To determine $\mu_\mathrm{MATBG}(n_\mathrm{MATBG})$, we track the $R_{xx}$ maxima along $\mu_\mathrm{MLG} = 0$, which is overlaid in light purple in Fig.~\ref{fig:chem_pot}\textbf{a}. We then use Supplementary equation 4 to determine $\mu_\mathrm{MATBG}(n_\mathrm{MATBG})$, which is shown in Fig.~1\textbf{g} of the main text. The corresponding change in $\mu_\mathrm{MATBG}(n_\mathrm{MATBG})$ across the flat bands is $\delta\mu \sim 40$ meV, comparable in magnitude to previous reports of MATBG~\cite{yu_spin_2022, park_flavour_2021, zondiner_cascade_2020, saito_isospin_2021, wong_cascade_2020}, further justifying the assumption of decoupled subsystems and the use of the bare graphene Fermi velocity.

\section{3. Landauer-Büttiker Analysis}\label{App:LB}

\begin{figure*}[t!]
    \renewcommand{\thefigure}{S\arabic{figure}}
    \centering
    \includegraphics[width=0.9\columnwidth]{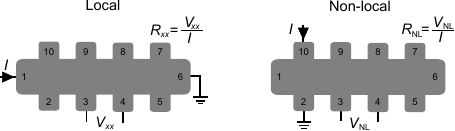}
    \caption{\textbf{Resistance measurement geometries}. Local (left) and non-local (right) resistance measurement geometries. Contacts are labeled 1-10.
} 
    \label{fig:geometry}
\end{figure*}

For a system where the transport is dominated by edge modes, the Landauer-Büttiker formula establishes the relationship between the edge mode transport and the measured resistance for arbitrary contact configurations. For the device studied here (Fig.~\ref{fig:geometry}) we consider a sample with $N$ edge modes coupled to 10 leads. Generically, the total transmission amplitude $T_{nm}$ between the $m$-th and $n$-th contacts of a sample is given by
\begin{equation}
T_{nm} = \sum_{\alpha, \beta = 1}^{N} |t_{m \alpha, n \beta}|^2,
\end{equation}
where $|t_{m \alpha, n \beta}|^2$ is the amplitude for the $\alpha$-th mode from terminal $m$ to tunnel into the $\beta$-th mode at terminal $n$. Here  $|t_{m \alpha, n \beta}|^2 = 1$ for perfect transmission and $|t_{m \alpha, n \beta}|^2 = 0$ for no transmission.

The Landauer-Büttiker formula relates the transmission amplitudes to the current at the $m$-th terminal, $I_m$, and the voltage at the $n$-th terminal, $V_n$, via
\begin{equation}
I_m = \frac{e^2}{h} (N V_m - \sum_{n} T_{mn} V_n),
\end{equation}
where the $e^2/h$ prefactor comes from the universal conductance of 1D channels. For perfect transmission in the case where the $N$ edge modes consist of $N_R$ right moving modes and $N_L$ left moving modes, we have $T_{m,m+1} = N_R$ and  $T_{m,m-1} = N_L$, while all other transmission amplitudes vanish. However, we can also consider the situation where the transmission is reduced by some amount of backscattering between the right and left moving edge modes. Assuming $N_R \geq N_L$ without loss of generality, we introduce a phenomenological parameter $0 \leq \tau \leq 1$ to parameterize the degree of backscattering. Physically, $\tau = 0$ indicates that there is no backscattering and $\tau = 1$ indicates complete backscattering, where the $N_L$ left moving modes fully scatter off the right moving modes, removing all left moving modes, and leaving $N_R - N_L$ right moving modes. There are special cases where rational values of the form $\tau = M/N_L \ (0 < M < N_L)$  can occur if $M$ of the $N_L$ left moving modes are fully backscattered by the right moving modes, leading to effectively $N_R - M$ right moving modes and $N_L - M$ left moving modes. In principle, such values of $\tau$ could also arise from a situation in which more than $M$ left moving modes partially scatter off the right moving modes, but obtaining those exact rational values of $\tau$ would require fine tuning. In general, other rational values and any irrational values of $\tau$ are necessarily fine tuned, and correspond to partial scattering of left and right moving modes.

\begin{table}[]
\renewcommand{\thetable}{S\arabic{table}}

    \centering
    \begin{tabular}{|p{1.3cm}|p{5cm}|p{5cm}|p{2.3cm}|p{2cm}|p{2.3cm}|}
    \hline
         & \boldmath$N_L = N_R = 1$ & \boldmath$N_L = N_R = 2$\\
    \hline
       \boldmath$\tau = 0$ & $R_{xx} = h/2e^2$, $R_\mathrm{NL} = h/5e^2$ & $R_{xx} = h/4e^2$, $R_\mathrm{NL} = h/10e^2$\\
    \hline
        \boldmath$\tau = 1/2$ & ----- & $R_{xx} = h/2e^2$, $R_\mathrm{NL} = h/5e^2$\\
    \hline
        \boldmath$\tau = 1$ & $R_{xx} = R_\mathrm{NL} = \infty$ & $R_{xx} = R_\mathrm{NL} = \infty$\\
    \hline

    \end{tabular}
    \caption{\textbf{Landauer-Büttiker formula predictions}. Predicted resistance values based on the Landauer-Büttiker formula (Supplementary equation 7) for different numbers of left/right propagating edge modes ($N_L$ and $N_R$) and scattering strengths $\tau$. We only list values corresponding to $\tau = M/N_R$, where $M \leq N_R$ is an integer.}
    \label{tab:table}
\end{table}

The phenomenological parameter $\tau$ enters the Landauer-Büttiker formula via the transmission amplitudes $T_{m,m+1} = N_R - \tau N_L$ and  $T_{m,m-1} = (1- \tau)N_L$. To calculate the expected $R_{xx}$, we substitute these transmission amplitudes into the Landauer-Büttiker formula (Supplementary equation 6). For our measurements we  set the source-lead current $I_1$ to $I_1 = I$, and the drain-lead current $I_6$ to $I_6 = -I$, and all other currents to zero (Fig.~\ref{fig:geometry}). The resistance $R_{xx} \equiv \Delta V / I$, where $\Delta V$ is the voltage difference across the probe leads 3 and 4, is found to be:
\begin{equation}
R_{xx} = \frac{h}{e^2} \frac{(N_R - \tau N_L)(1 - \tau) N_L}{(N_R - \tau N_L)^3+ ((1 - \tau) N_L)^3}.
\end{equation}
A similar derivation can be made for the non-local measurement geometries we use. For this geometry we set the source-lead current $I_{10}$ to $I_{10} = I$, and the drain-lead current $I_2$ to $I_2 = -I$, and all other currents to zero (Fig.~\ref{fig:geometry}). The resistance $R_\mathrm{NL} \equiv \Delta V / I$, where $\Delta V$ is the voltage difference across the probe leads 3 and 4, is
\begin{equation}
R_\mathrm{NL} = \frac{h}{e^2} \frac{N_L N_R^6}{N_R^8 + N_R^6 N_L^2 + N_R^4 N_L^4 + N_R^2 N_L^6  + N_L^8 }
\end{equation}
when $\tau = 0$. For $\tau > 0$, $R_\mathrm{NL}$ is related to the above calculation by substituting $N_R \rightarrow N_R - \tau N_L$ and $N_L \rightarrow (1 - \tau) N_L$. 

We next discuss the predicted values of $R_{xx}$ and $R_\mathrm{NL}$ for various cases directly relevant to what is studied in the main text. At $\nu_{\mathrm{MLG}}/\nu_{\mathrm{MATBG}} = \pm 1 / \mp 1$, we have $N_L = N_R = 1$ and Supplementary equations 7 and 8 simplify to:
\begin{equation}\begin{split}
& R_{xx} = \frac{h}{2e^2} \frac{1}{(1 - \tau)},\\
& R_\mathrm{NL} = \frac{h}{5e^2} \frac{1}{(1 - \tau)}.
\end{split}\end{equation}
If we assume $\tau$ can take only values $0$ or $1$, then either $R_{xx}$, $R_\mathrm{NL} = h/2e^{2}$, $h/5e^{2}$ or $R_{xx}$, $R_\mathrm{NL} \rightarrow \infty$, $\infty$, respectively. Our data are consistent with $\tau \sim 0$.

We are also interested in the case where $N_L = N_R = 2$, which is relevant for both the $\nu_{\mathrm{MLG}}/\nu_{\mathrm{MATBG}} = \pm 2 / \mp 2$ and $\nu_{\mathrm{MLG}}=\pm2$/$(t, s) = (\mp2, \mp2)$ states. The former correspond to quantum Hall states emanating from the charge neutrality point (CNP) in MATBG, and the latter to Chern insulators (ChIs) in MATBG, where $t$ is the Chern number and $s$ is the zero-field intercept. In this case Supplementary equations 7 and 8 become:
\begin{equation}\begin{split}
& R_{xx} = \frac{h}{4e^2} \frac{1}{(1 - \tau)},\\
& R_\mathrm{NL} = \frac{h}{10 e^2} \frac{1}{(1 - \tau)}.
\end{split}\end{equation}
If we now assume $\tau$ can take only the rational values $0$, $1/2$, or $1$, then either $R_{xx}$, $R_\mathrm{NL} = h/4e^{2}$, $h/10e^{2}$,  $R_{xx}$, $R_\mathrm{NL} = h/2e^{2}$, $h/5e^{2}$, or $R_{xx}$, $R_\mathrm{NL} \rightarrow \infty$, $\infty$  respectively. Our data for the $\nu_{\mathrm{MLG}}/\nu_{\mathrm{MATBG}} = \pm 2 / \mp 2$ and $\nu_{\mathrm{MLG}}=2$/$(t, s) = (-2, -2)$ state are most consistent with $\tau \sim 1$. In constrast, the $\nu_{\mathrm{MLG}}=-2$/$(t, s) = (2, 2)$ behavior is most consistent with $\tau \sim 1/2$, i.e., there is one pair of edge states that backscatter and another pair of counterpropagating modes that remain decoupled. In all cases, the transverse resistance $R_{xy}$ is zero. The predicted resistances for both $R_{xx}$ and $R_\mathrm{NL}$ for $N_L = N_R = 1, 2$ and $\tau = 0, 1/2$, or 1 are summarized in Table~\ref{tab:table}. 

\section{4. Edge state equilibration in a second contact pair}

\begin{figure*}[t!]
    \renewcommand{\thefigure}{S\arabic{figure}}
    \centering
    \includegraphics[width=0.7\columnwidth]{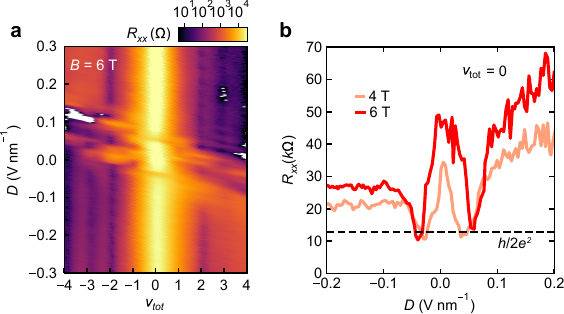}
    \caption{\textbf{Edge state equilibration near the CNP of a second contact pair}. \textbf{a}, $R_{xx}$ as a function of the total filling factor $\nu_{\mathrm{tot}} = \nu_{\mathrm{MLG}} + \nu_{\mathrm{MATBG}}$ and $D$ at $B = 6$ T near the CNP. \textbf{b}, $R_{xx}$ at $\nu_{\mathrm{tot}} = 0$ as a function of $D$ (extracted from \textbf{a}). The data agree closely with observations from the contact pair discussed in the main text.
} 
    \label{fig:2nd_contact_CNP}
\end{figure*}

\begin{figure*}[t!]
    \renewcommand{\thefigure}{S\arabic{figure}}
    \centering
    \includegraphics[width=1\columnwidth]{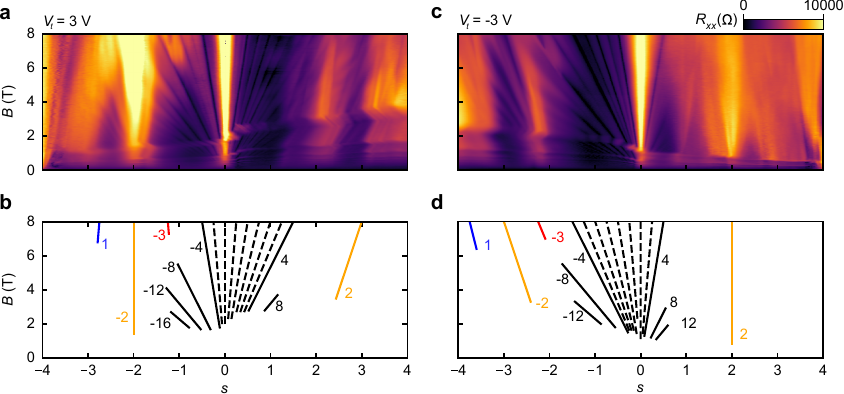}
    \caption{\textbf{Landau fan measurement of a second contact pair}. \textbf{a}, $R_{xx}$ as a function of $s$ and $B$ at fixed top gate voltage $V_t = 3$ V. \textbf{b}, Wannier diagram indicating the strongest quantum Hall and ChI states determined from \textbf{a}. The Chern numbers $t$ of the MATBG states are labeled. At high fields, the total Chern numbers of each state are offset by $2$ because $\nu_{\mathrm{MLG}} = 2$. Black, red, orange, and blue lines correspond to states with zero-field intercepts $s=0$, $s=|1|$,  $s=|2|$, and $s=|3|$, respectively. For states with $s=0$, $t\equiv\nu_\mathrm{MATBG}$. Black dashed lines label the MATBG symmetry broken quantum Hall states $-4 < \nu_\mathrm{MATBG} < 4$. \textbf{c}, \textbf{d}, Same as \textbf{a}, \textbf{b}, but for $V_t = -3$, where $\nu_{\mathrm{MLG}} = -2$ at high fields. Data collected at $T \approx 300$ mK.
}
    \label{fig:2nd_contact_fan}
\end{figure*}

\begin{figure*}[t!]
    \renewcommand{\thefigure}{S\arabic{figure}}
    \centering
    \includegraphics[width=1\columnwidth]{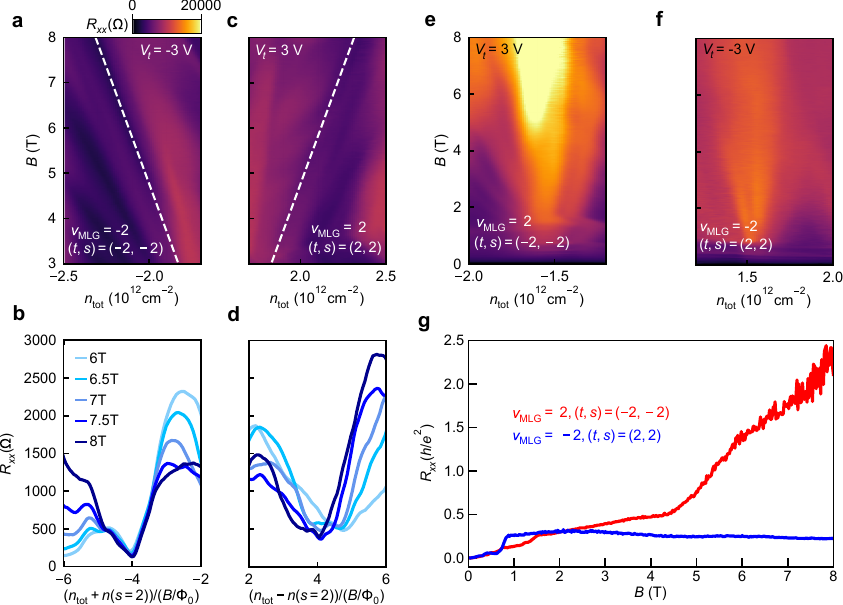}
    \caption{\textbf{Edge state equilibration of the even Chern insulators for a second contact pair}. \textbf{a}, $R_{xx}$ as a function of $n_\mathrm{tot}$ and $B$ for a second contact pair near the $(t, s) = (-2, -2)$ ChI in the co-propagating regime. \textbf{b}, $R_{xx}$ line cuts of the $\nu_{\mathrm{MLG}}=-2$/$(t, s) = (-2, -2)$ state at different $B$ forming well defined minima near zero. \textbf{c}, \textbf{d}, Same as \textbf{a}, \textbf{b}, but for the $\nu_{\mathrm{MLG}}=2$/$(t, s) = (2, 2)$ in the co-propogating regime. Minima in $R_{xx}$ are less well developed. \textbf{e}, \textbf{f}, $R_{xx}$ as a function of $n_\mathrm{tot}$ and $B$ for the second contact pair near the $(t, s) = (\mp 2, \mp2)$ ChIs, respectively, in the counter-propagating regime. \textbf{f}, $R_{xx}$ as a function of $B$ for the $\nu_{\mathrm{MLG}}=\pm2$/$(t, s) = (\mp2, \mp2)$ states. Data are identical to that presented in Fig.~\ref{fig:2nd_contact_fan}, but zoom-ins and line cuts of the relevant features are shown here for clarity. Data collected at $T \approx 300$ mK.
} 
    \label{fig:2nd_contact_Cherns}
\end{figure*}

In this section, we present additional measurements (Figs.~\ref{fig:2nd_contact_CNP} - \ref{fig:2nd_contact_Cherns}) from a second pair of contacts (2 and 3 in Fig.~\ref{fig:geometry}) that exhibit qualitatively similar behavior and therefore largely corroborate the conclusions about spin polarization discussed in the main text. For this second contact pair, $R_{xx}$ as a function of $\nu_\mathrm{tot}$ and $D$ at $B=6$ T (Fig.~\ref{fig:2nd_contact_CNP}\textbf{a}) matches closely to the observations from the contact pair discussed in the main text (3 and 4 in Fig.~\ref{fig:geometry}). Specifically, at $\nu_\mathrm{MLG}/\nu_\mathrm{MATBG} = \pm 1/ \mp 1$, local minima occur in $R_{xx}$ that are near the expected quantized value for a single pair of counter-propagating edge modes of opposite spin (Fig.~\ref{fig:2nd_contact_CNP}\textbf{b}).

Likewise, there is again an abrupt increase in $R_{xx}$ in this second pair as $|D|$ increases and the system enters the regime where $\nu_\mathrm{MLG}/\nu_\mathrm{MATBG} = \pm 2/ \mp 2$. The large $R_{xx}$ well above $h/2e^2$ indicates backscattering between both pairs of edge modes and therefore that the even-integer broken-symmetry quantum Hall states near the CNP in MATBG are spin unpolarized.

We next discuss the behavior of the ChIs in the second contact pair (Fig.~\ref{fig:2nd_contact_fan}), focusing first on the co-propagating regime when $\nu_{\mathrm{MLG}}=\pm2$. Similar to the contact pair studied in the main text, the $(t, s) = (-2, -2)$ ChI displays $R_{xx}$ minima near zero that follow a slope with a total Chern number of -4 (due to the contribution from $\nu_{\mathrm{MLG}}=-2$) in the co-propgating regime (Fig. \ref{fig:2nd_contact_Cherns}\textbf{a,b}). In contrast, the $(t, s) = (2, 2)$ ChI is less well formed in the co-propagating regime: the $R_{xx}$ minima are not as pronounced and their exact positions do not always perfectly fall along a single linear trajectory with a total Chern number of 4, as would be expected when accounting for the contribution from $\nu_{\mathrm{MLG}}=2$ (Fig. \ref{fig:2nd_contact_Cherns}\textbf{c,d}). 

In the counter-propagating regime with $\nu_{\mathrm{MLG}}=2$, the $(t, s) = (-2, -2)$ state displays high $R_{xx} > h/2e^2$ that increases with $B$, particularly above 5 T where the minima in $R_{xx}$ become well-developed (Fig. \ref{fig:2nd_contact_Cherns}\textbf{e,g}). This matches what is observed in the contact pair described in the main text, further supporting the conclusion that the $(t, s) = (-2, -2)$ ChI is spin unpolarized. In contrast, when $\nu_{\mathrm{MLG}}=-2$ the $(t, s) = (2, 2)$ ChI displays a much smaller $R_{xx}$ and weaker $B$-dependence (Fig. \ref{fig:2nd_contact_Cherns}\textbf{f,g}). The reduced $R_{xx}$ relative to the hole-doped side of MATBG is similar in some ways to the results presented in the main text. Importantly, however, the $R_{xx}$ of this contact pair is not quantized to $h/2e^2$; instead, the resistance is lower than $h/4e^2$, the expected value for two fully decoupled counter-propagating edge modes. This may in part reflect that the $(t, s) = (2, 2)$ ChI is itself less well formed, as is evident from its behavior in the co-propagating regime (Fig.~\ref{fig:2nd_contact_Cherns}\textbf{c},\textbf{d}). If the $(t, s) = (2, 2)$ state is not fully gapped or is more spatially inhomogeneous, this would likely lead to additional bulk contributions from the metallic-like character of the MATBG subsystem and therefore could ultimately lead to decreased resistance. For completeness, we also mention that it is possible that edge states with equal spin are not well coupled for this contact pair on the electron side, particularly given the generic observation of lower resistance for spin-unpolarized broken-symmetry states near the CNP when the MATBG is electron doped. However, we do not have an explanation for why this would be the case, nor why it would be more pronounced for this contact pair.

\section{5. Discussion of alternate explanations and quantitative resistance values}

In the main text, we highlight general features of the resistance observed in the counter-propagating edge mode regime and their most natural qualitative interpretations. In this section, we provide further quantitative discussion and address alternative potential explanations for the data in more detail.

\begin{figure*}[t!]
    \renewcommand{\thefigure}{S\arabic{figure}}
    \centering
    \includegraphics[width=0.7\columnwidth]{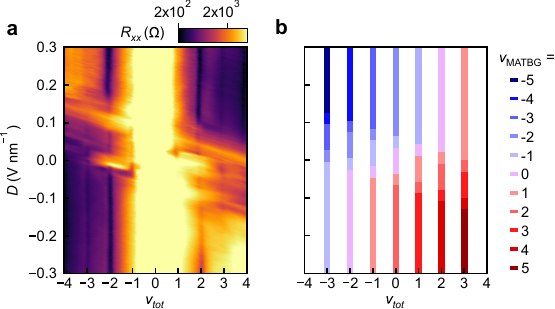}
    \caption{\textbf{Transport of the MATBG-MLG quantum Hall bilayer}. \textbf{a}, $R_{xx}$ as a function of $\nu_{\mathrm{tot}}$ and $D$ at $B = 8$ T near the CNP. The data are the same as presented in Fig. 2\textbf{a} in the main text, but are reproduced here for clarity. \textbf{b}, Schematic map of the possible combined MATBG and MLG quantum Hall filling factors. Each color represents a specific $\nu_{\mathrm{MATBG}}$ in the MATBG subsystem. Clear minima in $R_{xx}$ are, for example, observed at $\nu_{\mathrm{MLG}}/\nu_{\mathrm{MATBG}}=\pm2/\pm1$ and $\nu_{\mathrm{MLG}}/\nu_{\mathrm{MATBG}}=\pm1/0$, which are signatures of quantum Hall states $\nu_{\mathrm{MATBG}}=\pm1$ and $\nu_{\mathrm{MLG}}=\pm1$, respectively.
} 
    \label{fig:Rxx_coprop}
\end{figure*}

\subsection{a. Resistances at $\nu_{\mathrm{MLG}}/\nu_{\mathrm{MATBG}}=\pm1/\mp1$}

We focus first on the behavior at $\nu_\mathrm{MLG}/\nu_\mathrm{MATBG} = \pm 1/ \mp 1$. The observed longitudinal and non-local resistances are close to their expected quantized values for a pair of decoupled counter-propagating edge modes, but they do exhibit deviations toward lower resistance, especially at the highest magnetic fields. Below, we discuss potential causes for such discrepancies and why the interpretation of opposite-spin counter-propagating edge states remains the most natural explanation for the data.

There is evidence of quantum Hall states and edge mode transport in both MATBG and MLG subsystems at the relevant fillings. For example, we observe local minima in $R_{xx}$ in the co-propagating regime when $\nu_{\mathrm{MLG}}/\nu_{\mathrm{MATBG}}=\pm2/\pm1$, both in measurements at fixed $B$ (Fig.~\ref{fig:Rxx_coprop} and Fig.~\ref{fig:2nd_contact_CNP}\textbf{a}) and in Landau fans at fixed $V_t$ (Fig.~3 and Fig.~\ref{fig:2nd_contact_fan}). We also observe local minima in $R_{xx}$ when $\nu_{\mathrm{MLG}}/\nu_{\mathrm{MATBG}}=\pm1/0$ (Fig.~\ref{fig:Rxx_coprop}). These observations indicate developing MATBG and MLG quantum Hall states at $\nu_{\mathrm{MATBG}} = \pm1$ and $\nu_{\mathrm{MLG}} = \pm1$. Furthermore, as discussed in the main text, the appreciable $R_{\mathrm{NL}}$ at $\nu_\mathrm{MLG}/\nu_\mathrm{MATBG} = \pm 1/ \mp 1$ demonstrates the presence of counter-propagating edge modes. Likewise, we observe steps in chemical potential which provide additional thermodynamic evidence for a bulk gap in the MLG subsystem at $\nu_{\mathrm{MLG}} = \pm1$. These manifest as features in $R_{xx}$ that persist at constant filling over a range in displacement field; their extent in $D$ is a proxy for the size of the gap at $\nu_{\mathrm{MLG}}=\pm1$. Well-formed features are evident at $\nu_{\mathrm{MLG}} = \pm1$ (see, e.g., Fig.~\ref{fig:2nd_contact_CNP} and Fig.~\ref{fig:Rxx_coprop}).

Despite the above signatures consistent with quantum Hall states at the relevant fillings, we do not measure perfectly quantized resistance. One way that $R_{xx} < h/2e^2$ could arise is if there is some parallel bulk conduction (e.g. if the quantum Hall states are not fully developed or if the device is inhomogeneous). We note that at $\nu_\mathrm{MLG}/\nu_\mathrm{MATBG} = 0/0$, $R_{xx}$ is on the order of 40 - 60 k$\Omega$, which can be used as an estimate of residual bulk conduction because no edge modes are expected there. A resistance of 40 - 60 k$\Omega$ in parallel with the expected $h/2e^2 = 12.9$ k$\Omega$ edge mode resistance would give a total resistance of 9.8 - 10.6 k$\Omega$. These values are consistent with those that we observe, suggesting that additional weak bulk conduction could reasonably explain the small deviations in $R_{xx}$ below $h/2e^2$. 

Effects of parallel bulk conduction between contacts should be avoidable by measuring resistance in a non-local geometry, which is sensitive only to edge contributions. Indeed, at $\nu_\mathrm{MLG}/\nu_\mathrm{MATBG} = -1/ 1$, we observe that $R_\mathrm{NL}$ exhibits a plateau-like feature very close to $h/5e^2$,  the predicted quantized value for this geometry. However, at $\nu_\mathrm{MLG}/\nu_\mathrm{MATBG} = 1/ -1$, $R_\mathrm{NL}$ drops below $h/5e^2$. One likely contributing factor for this dip in $R_\mathrm{NL}$ is an artifact due to regions of the contacts to the device that are top gated but not bottom gated. Specifically, we observe a diagonal feature with negative $R_\mathrm{NL}$ that is approximately parallel to the bottom gate voltage $V_b$ axis (Fig.~\ref{fig:non_local}) and intersects the $\nu_\mathrm{MLG}/\nu_\mathrm{MATBG} = 1/ -1$ state (but \textit{not} the $\nu_\mathrm{MLG}/\nu_\mathrm{MATBG} = -1/ 1$ state). The correlation of this feature with $V_b$ suggests that it arises in a portion of the contact that is not bottom gated. This may spuriously lower the measured $R_{\mathrm{NL}}$ of the $\nu_\mathrm{MLG}/\nu_\mathrm{MATBG} = 1/ -1$ state. Additionally, it is possible the lower $R_\mathrm{NL}$ may in part be due to `leakage' of the edge current into alternate current paths through the bulk that do not connect the probed pair of contacts, i.e. that terminate at a different contact.

Given the above evidence for incipient quantum Hall states at the relevant fillings, signatures of edge conductance in non-local resistance, and measured resistance which approaches the quantized values expected for decoupled counter-propagating edge modes (and potential explanations for deviations) for multiple contact pairs, we conclude that the most likely explanation is that the MLG and MATBG states have opposite spin polarization at $\nu_\mathrm{MLG}/\nu_\mathrm{MATBG} = \pm 1/ \mp 1$. This is also consistent with expectations for spin polarization based on the Zeeman effect.

\subsection{b. Resistances at $\nu_{\mathrm{MLG}}/\nu_{\mathrm{MATBG}}=\pm2/\mp2$ and $\nu_{\mathrm{MLG}}=2 / (t,s)=(-2,-2)$}

In the counter-propagating regime where $\nu_{\mathrm{MLG}}/\nu_{\mathrm{MATBG}}=\pm2/\mp2$, as well as when $\nu_{\mathrm{MLG}}=2$ and $(t,s)=(-2,-2)$, the measured $R_{xx}$ and $R_{\mathrm{NL}}$ both exceed the corresponding quantized values expected for a single decoupled pair of counter-propagating edge modes. As noted in the main text, this indicates that there is backscattering between both pairs of counter-propagating modes and therefore that the MATBG state is spin-unpolarized in all of those cases. However, especially when the MATBG is electron doped, the measured resistances do not exceed the quantized values by a very large factor. Therefore, we discuss alternative potential explanations for the data below.

One possibility to consider is whether sample inhomogeneity could lead to disorder-induced equilibration of edge modes even if they have opposite spins. If there are multiple segments (separated by disordered regions) between contacts, each with a quantized resistance, they would add in series and could increase the measured resistance beyond a quantized value~\cite{Wu_2018, Essert_2015, Maciejko_2010}. We view this scenario as highly unlikely for several reasons. First, there is relatively low twist angle disorder (Fig.~\ref{fig:contacts}), and we observe qualitatively similar behavior for multiple contact pairs as well as in non-local measurements. Second, although the maximum resistance only reaches a few times the quantized value, it is an appreciable fraction of the resistance at $n_{\mathrm{tot}}=0$ and $D=0$, where both MLG and MATBG are in broken-symmetry states at $\nu=0$, and no bulk conductance or edge modes are expected. Finally, we observe close to quantized values when $\nu_\mathrm{MLG}/\nu_\mathrm{MATBG} = \pm 1/ \mp 1$, indicating a decoupled pair of edge modes. There is no obvious reason to expect that these would not also be similarly affected by disorder along the edge. Thus, the abrupt jump in resistance from the $\nu_\mathrm{MLG}/\nu_\mathrm{MATBG} = \pm 1/ \mp 1$ states to the $\nu_\mathrm{MLG}/\nu_\mathrm{MATBG} = \pm 2/ \mp 2$ states provides strong evidence for significantly enhanced backscattering between the edge modes in the latter case.

It is also worth commenting on the fact that $R_{xx}$ is lower for the $\nu_\mathrm{MLG}/\nu_\mathrm{MATBG} =  2/ - 2$ state than the  $\nu_\mathrm{MLG}/\nu_\mathrm{MATBG} = - 2/ 2$ state, particularly when measured using contacts 3 and 4 (the primary contact pair studied in the  main text). Despite this difference, our data still suggest that both topological sectors host two pairs of counter-propagating spin unpolarized edge modes and that the edge modes can efficiently back scatter off each other. This conclusion is motivated by the fact that the $R_{\mathrm{NL}}$ of the $\nu_\mathrm{MLG}/\nu_\mathrm{MATBG} =  2/ - 2$ state (which should be primarily determined by edge mode transport) is more than twice as large as the expected $R_\mathrm{NL}$ of a system with a single counter-propagating pair of edge modes ($h/5e^2$ in this geometry). Additionally, the value of  $R_{\mathrm{NL}}$ is comparable between the $\nu_\mathrm{MLG}/\nu_\mathrm{MATBG} = 2/-2$ and $\nu_\mathrm{MLG}/\nu_\mathrm{MATBG} = -2/2$ state. This indicates that the relatively lower $R_{xx}$ of the $\nu_\mathrm{MLG}/\nu_\mathrm{MATBG} =  2/ - 2$ state is likely due to weak parallel bulk conduction, possibly caused by a difference in the size of the gaps of these states due to electron-hole asymmetry in MATBG.
We additionally note that the $R_{xx}$ for the second contact pair discussed in Section 3 displays $R_{xx} > h/e^2$ for both $\nu_\mathrm{MLG}/\nu_\mathrm{MATBG} = \pm2/\mp2$, and better quantized  $R_{xx} = h/2e^2$ at $\nu_\mathrm{MLG}/\nu_\mathrm{MATBG} = \pm1/\mp1$, suggesting that this region of the device may be less susceptible to additional bulk contributions compared to the primary contact pair. Finally, recent calculations indicate the possibility of partial spin polarization of the broken-symmetry states due to interaction-driven renormalization of Hofstadter subbands~\cite{wang2023theory}. If there is electron-hole asymmetry in the MATBG band structure and/or exchange interaction, this may affect the relative degree of spin polarization of the two states and therefore the inter-edge backscattering and measured resistances.

 \subsection{c. Resistance at $\nu_{\mathrm{MLG}}=-2$ and $(t, s) = (2, 2)$}

Finally, we discuss in more detail alternate scenarios that could lead to the observed resistances of the $\nu_{\mathrm{MLG}}=-2$ and $(t, s) = (2, 2)$ state. The longitudinal and non-local resistances of this state exhibit plateaus that are nearly quantized at the values predicted by the Landauer-Büttiker formalism (Supplementary equation 9) for a single pair of counter-propagating edge modes. This is consistent with a picture where the state initially has two pairs of counter propagating edge modes, but efficient backscattering occurs only for one pair, resulting in a topological state having a single remaining pair of counter propagating edge modes (see Sec. 2 for further discussion). This would naturally occur if the $(t, s) = (2, 2)$ is spin-polarized: when combined with the spin-unpolarized edge modes of the $\nu_{\mathrm{MLG}}=-2$ state, this leads to a pair of counter-propagating modes having the same spin (that can efficiently backscatter) and one pair of counter-propagating modes having opposite spin that \textit{cannot} efficiently scatter off each other (in the absence of magnetic impurities). 

However, other mechanisms for decoupling are possible and could contribute to the lower resistances we observe (in all contact pairs) relative to the $\nu_{\mathrm{MLG}}=2$/$(t, s) = (-2, -2)$ state. We discuss these possibilities below and the degree to which they are likely to be contributing factors.

One possibility to consider is that both the $\nu_{\mathrm{MLG}}=-2$ and $(t, s) = (2, 2)$ edge modes are spin unpolarized, but with weakened symmetry-allowed scattering processes. This is unlikely because such scattering processes appear to be robust for the other states we measure. The near-quantization of both $R_{xx}$ and $R_{\mathrm{NL}}$ for the contacts discussed in the main text would also require fine-tuning, making this explanation less likely (though not impossible). This contrasts with the natural expectation for a spin-polarized state.

Another possibility is that the $(t, s) = (2, 2)$ is not well-developed, similar to what is observed for the second contact pair discussed above in Supplementary Note 4 of the supplement. This explanation, however, is not consistent with the well-developed zeros in $R_{xx}$ and quantized $R_{xy} = h/4e^2$ in the co-propagating regime for the primary pair contacts (Fig.~\ref{fig:Landau_Chern_Zoom_Co}). For these reasons, the $R_{xx}$ and $R_\mathrm{NL}$ for the contact pair discussed in the main text suggests that the $(t, s) = (2, 2)$ ChI is most likely spin polarized.

\section{6. Interpretation of results in the Hofstadter subband picture}

\begin{figure*}[t!]
    \renewcommand{\thefigure}{S\arabic{figure}}
    \centering
    \includegraphics[width=1\columnwidth]{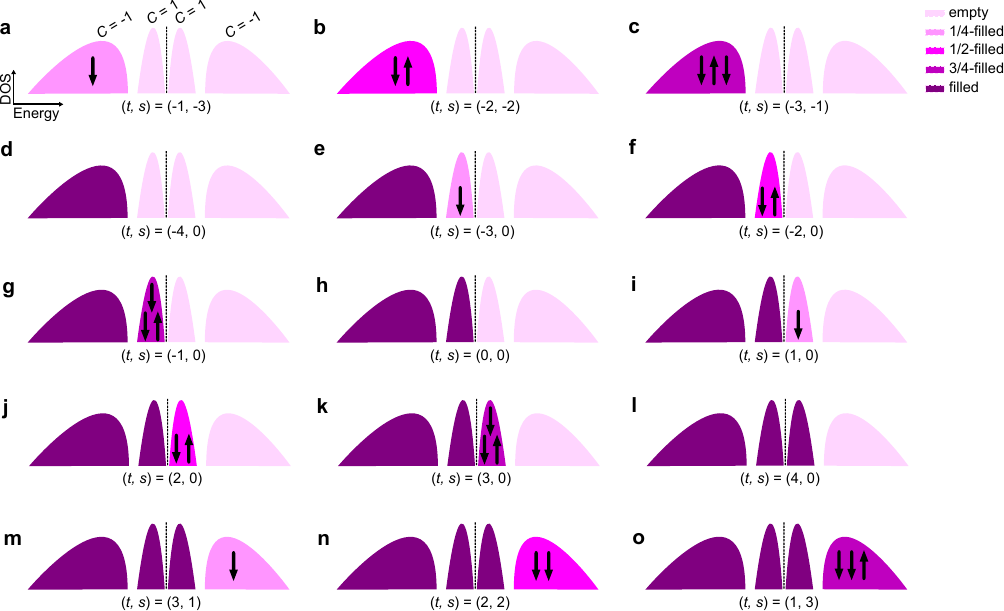}
    \caption{\textbf{Spin filling sequence of the Hofstadter subbands}. Schematic representation of the Hofstadter subbands as they are filled with different spin states. The Chern numbers of each subband are labeled in the top left image, and the central dashed line marks the charge neutrality point. Note that these diagrams are meant only as a cartoon representation and that the relative heights and bandwidths do not reflect calculated values. The different colors represent different levels of filling for each of the subbands, as indicated in the legend. The arrows represent the inferred spin polarization of the flavors occupying the bands.
} 
    \label{fig:subbands}
\end{figure*}

In this section, we discuss the inferred sequence of fillings of Hofstadter subbands based on our data.
In a perpendicular magnetic field, MATBG states form a fractal Hofstadter energy spectrum~\cite{Hofstadter_1976, Bistritzer_2011, Moon_2012, Hejazi_2019, Zhang_2019, Lian_2020, Wang_2022, Herzog-Arbeitman_2022, parker2021fieldtuned, wang2023theory}, which is relevant to both the ChIs and the quantum Hall states emanating from the CNP.
The spectrum is smeared out by temperature and disorder so that only the largest gaps between subbands are experimentally resolved~\cite{saito_hofstadter_2021, yu_correlated_2022, park_flavour_2021, choi_correlation-driven_2021}. This leads to three especially prominent subbands with respective Chern numbers of $-1$, $2$, and $-1$; there are four copies of this spectrum due to the four possible flavors from the spin and valley degrees of freedom in MATBG. 
Furthermore, the central $C = 2$ subband arises from the two Dirac crossings (per spin and valley flavor) at the MATBG CNP, i.e. the two moiré valleys.
This central subband can be further split into two $C = 1$ subbands in the presence of moiré valley splitting, which can arise from heterostrain and/or interactions that break $M_y$ symmetry. As discussed below, we observe evidence of such splitting and thus schematically illustrate a central doublet of $C=1$ subbands in Fig.~\ref{fig:subbands}. The primary sequence of ChIs in MATBG (satisfying $|t+s|=4$) is well described by sequentially filling the $C = -1$ bands with different particle flavors, while the quantum Hall states near the charge neutrality arise from a sequential filling of the two central $C = 1$ subbands. The total Chern number $t$ of a given state in MATBG is the sum of the Chern numbers of all the filled subbands.

For topological phases of matter, the bulk-edge correspondence predicts that the net number of low-energy topologically protected edge states matches the bulk Chern number. Thus, like quantum Hall states, ChIs arising from occupation of Hofstadter subbands are also expected to exhibit an insulating bulk and chiral edge modes. As discussed in the main text, our experiments allow us to infer the spin polarization of these MATBG edge modes for both the quantum Hall states near the CNP and the ChIs. We can then infer the spin-resolved filling sequence of the topological Hofstadter subbands in MATBG from our data, which is schematically illustrated in Fig.~\ref{fig:subbands}.

As detailed in the main text, the data indicate that the spin polarization of the edge mode of $(t, s) = (-1, -3)$ is identical to that of $\nu_{\mathrm{MLG}}=1$, which corresponds to just a single flavor occupying the lowest-energy $C = -1$ subband. Thus in Fig.~\ref{fig:subbands}\textbf{a}, we show the $(t, s) = (-1, -3)$ with a single spin-down electron occupying the the first $C = -1$ subband. For $(t, s) = (-2, -2)$, we observe the edge modes to be spin-unpolarized. Thus the next flavor to occupy the first $C = -1$ subband must be spin-up (Fig.~\ref{fig:subbands}\textbf{b}). This, together with Zeeman considerations, also allows us to deduce the spin polarization in the (-3,-1) and (-4,0) states (Fig.~\ref{fig:subbands}\textbf{c},\textbf{d}), the latter of which is a flavor symmetric state (all four flavors filled). 

Once the first $C = -1$ subband is fully filled (with 2 spin-down and 2 spin-up flavors), the central $C = 1$ subbands begin to be occupied. At $\nu_{\mathrm{MATBG}} = -2$ [equivalent to $(t, s)=(-2, 0)$], our data indicates spin unpolarized edge modes, which corresponds to one spin-up and spin-down flavor occupying the first $C = 1$ subband (Fig.~\ref{fig:subbands}\textbf{f}). Note that if the central subband was \textit{not} split and instead retained its $C = 2$ Chern number, then exchange interactions would favor occupying it with a single flavor, which would lead to a spin-polarized state at $\nu_{\mathrm{MATBG}} = -2$. That we instead observe a spin-unpolarized state is evidence of moiré valley splitting and the formation of a $C=1$ doublet of subbands. The same argument applies for the $\nu_{\mathrm{MATBG}} = 2$ state [equivalent to $(t, s)=(2, 0)$], where we also observe spin-unpolarized edge modes. Together with the spin identification based on counter-propagating opposite-spin states at $\nu_\mathrm{MLG}/\nu_\mathrm{MATBG} = \pm1/\mp1$ [equivalent to $(t,s)=(\pm1,0)$], and assuming the Zeeman effect dictates the spin of the $(\pm3,0)$ states, we determine the spin filling sequence of the central Hofstadter subbands in Fig.~\ref{fig:subbands}\textbf{e}-\textbf{l}.

Once the first three subbands are all filled, electrons begin filling the highest energy $C = -1$ subbands to form the ChIs on the electron side. Our data suggests that the $(t, s) = (2, 2)$ state is likely spin-polarized. Thus we tentatively illustrate this state with two spin-down electrons in the highest energy $C = -1$ subbands (spin-down assuming the Zeeman effect favors this over spin-up) (Fig.~\ref{fig:subbands}\textbf{n}). Note that although two additional spin-down flavors occupy the subband, the edge modes will be spin-up. This is because the additional two $C = -1$ spin-down edge modes will effectively ``cancel" out the two $C = 1$ spin-down edge modes from the $(t, s) = (4, 0)$ state, leaving behind two remaining spin-up edge modes in the $(t, s) = (2, 2)$ state. This reasoning leads to the filling sequence for the uppermost $C=-1$ subband illustrated in (Fig.~\ref{fig:subbands}\textbf{m}-\textbf{o}).

\section{7. Discussion of generality of the results}

For most combinations of MLG/MATBG states, our data provide consistent results from multiple contact pairs and in measurements conducted in a non-local geometry. Furthermore, the features in the data that we ascribe to MATBG match well to previously observed generic behavior in MATBG devices. However, we acknowledge that the measurements presented here are primarily from one device, and it is possible that the spin polarization of correlated states could have some sample dependency. This is common in moiré systems such as MATBG, where microscopic details, such as strain and twist angle, can lead to different observed correlated ground states~\cite{nuckolls2023quantum, kwan_2021, Wagner_2022, Liu_2021, parker_2021, Bi_2019}. 
Even if such microscopic details modify device specific behavior, our work clearly demonstrates that symmetry breaking terms arising from exchange effects and strain are competitive with Zeeman splitting, and that even the preference for spin-polarized versus unpolarized ground states can exhibit electron-hole asymmetry.
Importantly, the measurements also serve as a proof of principle demonstration of a new device architecture and a powerful approach to identify quantum degrees of freedom, applied to a sample whose features are consistent with prior reports in MATBG.

\section{Supplementary References}

\bibliographystyle{apsrev4-1}
\bibliography{references.bib}